\documentclass[aps,pre,twocolumn,superscriptaddress,tight]{revtex4-1}

\usepackage{amssymb}
\usepackage{amsfonts}
\usepackage{color}
\usepackage{graphicx}
\usepackage{amsmath}
\usepackage{times}
\usepackage{bm}
\usepackage{float}
\usepackage{lipsum}
\usepackage{import}
\usepackage{enumerate}
\usepackage[colorlinks,linkcolor=blue,anchorcolor=blue,citecolor=blue]{hyperref}

\begin{document}

\title{A Unified Ott--Antonsen Framework for Stability and Bifurcations in Ring Swarmalators}
		
\author{Can Xu}
\affiliation{Institute of Systems Science and College of Information Science and Engineering, Huaqiao University, Xiamen 361021, China}

\author{Wenhan Xu}
\affiliation{School of Mathematical Sciences, Huaqiao University, Quanzhou 362021, China}
	
\author{Xingang Wang}
\email{Contact author: wangxg@snnu.edu.cn}
\affiliation{School of Physics and Information Technology, Shaanxi Normal University, Xi'an 710062, China}
\begin{abstract}
Swarmalators, a prototypical model that integrates temporal rhythms with spatial swarming, have been widely used to explore the collective dynamics observed in active-matter systems. Although numerical simulations have revealed a variety of collective states, a general theoretical framework for the stability and bifurcations of the states remains unavailable. Here, by the model of one-dimensional ring-structured swarmalators with general distributions of natural frequencies and velocities, we conduct a systematic theoretical analysis of the stability and bifurcations of four representative collective states observed in simulations: the desynchronization state, the phase-wave state, the synchronization state, and the mixed state. We show that the macroscopic dynamics are governed by a low-dimensional invariant manifold, namely the generalized Ott--Antonsen manifold, and further establish that the same manifold underlies the collective dynamics of high-dimensional agent models, thereby unifying these seemingly different descriptions within a common reduction framework. Building on this connection, we derive stability conditions and bifurcation boundaries for the collective states and construct a bifurcation diagram in the two-parameter coupling plane, with theoretical predictions validated by large-scale numerical simulations. Our results clarify the dynamical origins of state selection in swarmalator systems and provide a theoretical foundation for analyzing nonequilibrium phase transitions in active-matter systems with coupled internal rhythms and spatial motions.\\ \\
\noindent{\bf Subject Areas:} Swarmalators, Collective states, Ott--Antonsen manifold, Bifurcation diagram
\end{abstract}

\maketitle

\section{Introduction}\label{Sec I}

Synchronization is ubiquitous in natural and man-made complex systems~\cite{1,2,3}, with examples ranging from the flashing of fireflies to the contraction of cardiac cells, and from biological circadian rhythms to power grids. A standard approach to modeling synchronization in populations of coupled dynamical units is to embed the units in a \emph{topological} space: each unit is represented by a node, and interactions are encoded by a network describing who couples to whom. In this setting, geometric (spatial) properties of the units and their interactions are neglected. To incorporate geometric information while retaining a network description, weighted networks have been proposed to model the dynamics of spatially distributed systems~\cite{4,5}, with coupling weights encoding spatial effects. Such weighted-network models, however, are most suitable for systems with immobile units (e.g., power grids and neuronal networks) and generally fail to capture complex systems composed of \emph{mobile} units (e.g., bird flocks, bee swarms, and fish schools). In mobile systems, the intrinsic dynamics (the evolution of the units in phase space) and the extrinsic dynamics (their motion in geometric space) are mutually coupled, leading to novel spatiotemporal phenomena absent in systems of immobile units~\cite{6,7}.

Motivated by recent advances in multi-agent systems, particularly active and soft matter, the collective behaviors of mobile dynamical units have attracted increasing attention. For such systems to function, it is often necessary not only that intrinsic states become correlated, but also that motion in geometric space becomes coordinated. A paradigmatic example is the model of \emph{swarmalators}, where each unit carries an oscillator as its intrinsic dynamics while simultaneously swarming in space. As a general framework integrating spatial swarming and temporal synchronization, swarmalators are relevant to a broad class of natural and artificial systems, including Japanese tree frogs~\cite{8}, the nematode \emph{Turbatrix aceti}~\cite{9}, magnetotactic bacteria~\cite{10}, magnetic domain walls~\cite{11}, Quincke rotators~\cite{12}, and micro- or nano-scale colloidal particles such as Janus particles~\cite{13}. These broad implications have made swarmalator dynamics an active topic in both experimental and theoretical studies.

Although synchronization and swarming have each been extensively studied, their interplay, as embodied by swarmalators, remains an open question. Classical synchronization models, such as the Stuart--Landau model of coupled limit-cycle oscillators~\cite{14,15} and the Kuramoto model of coupled phase oscillators~\cite{16,17,18}, aim to explain the emergence of synchrony from the viewpoints of nonlinear dynamics and nonequilibrium phase transitions~\cite{19,20,21,22,23,24}, with the central task being determining the critical coupling for the synchronization transition. By contrast, canonical swarming models, such as the Vicsek model~\cite{25}, focus on the coordination of self-propelled particles in geometric space, where the key issue is finding the conditions for alignment (e.g., speed, density, and noise intensity). While both classes of models address collective behaviors in interacting ensembles~\cite{26,27,28,29}, they do so from fundamentally different perspectives: temporal coherence of intrinsic oscillator states versus spatial alignment of mobile agents. Accordingly, the theoretical tools for analyzing synchronization differ substantially from those developed for swarming. As a hybrid model integrating both ingredients, swarmalator dynamics is therefore particularly challenging to analyze~\cite{30,31}.

A brief review of the theoretical progress on swarmalator systems is as follows. Early foundations were laid by Tanaka and Isawa~\cite{32,33}, who modeled and analyzed the collective dynamics of chemotactic oscillators. These active systems exhibit both chemotaxis and autonomous oscillations and occur in microorganisms such as \emph{Escherichia coli} and in synthetic microswimmers. In their framework, oscillators move in geometric space according to local chemical gradients; the chemical field is, in turn, modulated by the intrinsic oscillator states, while the oscillator dynamics is influenced by the chemical concentration. By the methods of center-manifold and phase-reduction, they derived a reduced model and showed that by varying the system parameters, the system can generate diverse spatiotemporal patterns, e.g., spatially distributed clusters with antiphase synchronization. A major breakthrough was achieved by O'Keeffe {\it et al.} in 2017, who introduced a two-dimensional swarmalator model in which spatial motion is coupled to Kuramoto-like phase dynamics. In this model, spatial movement depends on the intrinsic phases, while the coupling among units depends on their pairwise spatial distances. Despite its simple form, the model exhibits remarkably rich collective dynamics, including synchronization, desynchronization, and various phase-wave states. This heuristic framework has since been extended in multiple directions, and the typical collective states have been observed in experimental and natural systems, motivating continued investigation~\cite{35,36,37,38,39}.

Despite extensive numerical evidence, the theoretical analysis of the two-dimensional swarmalator model remains difficult. Methods such as renormalization group theory and nonequilibrium field theory are powerful for mobile-agent models without intrinsic dynamics (e.g., the Vicsek model~\cite{40}), but they are generally not directly applicable to active systems with intrinsic dynamics, such as swarmalators, because the relevant order parameters, collective behaviors, and bifurcation structures differ qualitatively. To address this challenge, Ref.~\cite{41} introduced a one-dimensional solvable model of nonidentical swarmalators and reported rich collective states. Remarkably, these states can be analyzed using the Ott--Antonsen ansatz from synchronization theory, yielding both their spatiotemporal properties and the conditions for their existence. This one-dimensional solvable model has opened a new avenue for studying swarmalator systems and has been widely used to investigate collective dynamics in active-matter settings~\cite{42,43,44,45,46,47,48,49,50,51}.

In Ref.~\cite{41}, the authors first proposed a one-dimensional solvable swarmalator model, reported four prototypical spatiotemporal collective states, and provided qualitative and semi-quantitative explanations for pattern emergence under specific frequency-velocity distributions using the Ott-Antonsen ansatz, thereby offering crucial theoretical insights for a series of subsequent studies. Nevertheless, three fundamental issues remain to be resolved: Why does the OA ansatz, as a low-dimensional invariant manifold, apply to ring-shaped swarmalator systems? What are its universality and its intrinsic connection to high-dimensional agents? Under general frequency-velocity distributions, can the instability threshold of the desynchronization state be described by a unified expression that integrates frequency and spatial disorder? And what are the eigenspectral structures and bifurcation mechanisms of other spatiotemporal collective states under general frequency-velocity distributions? The primary goal of our present work is to provide a systematic and in-depth investigation of these issues.

Motivated by the aforementioned issues, we revisit the one-dimensional swarmalator model proposed in Ref.~\cite{41} and investigate its connection to the high-dimensional agent model introduced by Tanaka in Ref.~\cite{52}. We are able to argue mathematically and demonstrate numerically that, despite their different dimensions, the two models are governed by the same low-dimensional invariant manifold, thereby unifying them within a single framework. Building on this connection, we study the dynamical origins of four types of collective states in swarmalator systems (the desynchronization state, the phase-wave state, the synchronization state, and mixed states) under general distributions of natural frequencies and velocities. Specifically, we characterize the conditions for their emergence, their stability, the associated eigen-spectrum features (in both discrete and continuous cases), and the mechanisms underlying the bifurcations. These results provide a general framework for analyzing the collective dynamics in swarmalators and lay a theoretical foundation for broader investigations of active-matter systems.

The remainder of this paper is organized as follows. In Sec.~\ref{Sec II}, we shall introduce the one-dimensional ring-structured swarmalator model and report collective states observed in simulations. In Sec.~\ref{Sec III}, we shall analyze the stability of the desynchronization state using a continuity-equation approach and derive the critical coupling strength at which it loses stability under general conditions. In Sec.~\ref{Sec IV}, we shall present the theoretical framework for analyzing the collective states and clarify the connection between the one-dimensional swarmalator model and the high-dimensional agent model from the perspective of the generalized Ott--Antonsen manifold. In Sec.~\ref{Sec V}, we shall apply the new framework to revisit the collective states, including the stability of the desynchronization state, the conditions for generating the phase-wave, synchronization, and mixed states, and the structures of their eigen-spectra. Finally, concluding remarks and potential directions for future research are given in Sec.~\ref{Sec VI}.

\section{Model and Phenomena}\label{Sec II}

Following Ref.~\cite{41}, we consider an ensemble of swarmalators moving on a one-dimensional ring. The dynamic of the swarmalators are governed by the equations
\begin{subequations}\label{eq:KM1}
\begin{align}
\dot{\theta}_i=\omega_i+\frac{K}{N}\sum_{j=1}^{N}\sin{\left(\theta_j-\theta_i\right)}\cos{\left(x_j-x_i\right)},\label{eq:KM1a} \\
\dot{x}_i=v_i+\frac{J}{N}\sum_{j=1}^{N}\sin{\left(x_j-x_i\right)}\cos{\left(\theta_j-\theta_i\right)}.\label{eq:KM1b}
\end{align}
\end{subequations}
Here, $\theta_i(t)$ and $x_i(t)$ denote the instantaneous phase and spatial coordinate of the $i$th swarmalator, respectively. The index $i$ runs from $1$ to $N$, where $N$ is the system size. In the uncoupled limit, the intrinsic dynamics in phase space is characterized by the natural frequency $\omega_i$, and the spatial dynamics in geometric space is characterized by the natural velocity $v_i$; both $\omega_i$ and $v_i$ are drawn independently from prescribed distributions. The interaction strengths are controlled by $K\geq 0$ and $J\geq 0$ for the phase and spatial dynamics, respectively.

Equation~(\ref{eq:KM1a}) is generalized from the classical Kuramoto model, as the effective coupling between swarmalators $i$ and $j$ in phase space is $K\cos(x_j-x_i)$. Thus, the coupling received by the $i$th swarmalator depends on its spatial distance to other swarmalators, providing a direct mechanism through which spatial motion modulates phase synchronization. Likewise, Eq.~(\ref{eq:KM1b}) has the same structure as Eq.~(\ref{eq:KM1a}) with $\theta_i$ replaced by $x_i$. The effective interaction in spatial space is $J\cos(\theta_j-\theta_i)$, so swarmalators with closer phases interact more strongly in space. This term couples the intrinsic dynamics back to the spatial dynamics.

Intuitively, Eqs.~(\ref{eq:KM1a}) and (\ref{eq:KM1b}) can be viewed as two coupled Kuramoto-type models. Interpreting swarmalators as agents moving on the surface of a torus, Eq.~(\ref{eq:KM1a}) describes motion along the meridional direction, while Eq.~(\ref{eq:KM1b}) describes motion along the circumferential direction. Unlike recently studied high-dimensional Kuramoto models, which focus solely on intrinsic oscillator dynamics, Eqs.~(\ref{eq:KM1a}) and (\ref{eq:KM1b}) capture spatiotemporal patterns self-organized by the interplay between phase dynamics (temporal rhythms) and spatial dynamics (swarming behaviors). This feature makes the model a suitable framework for exploring collective dynamics in active matter.

Introducing the variables $\zeta_i=x_i+\theta_i$ and $\eta_i=x_i-\theta_i$, Eqs.~(\ref{eq:KM1a}) and (\ref{eq:KM1b}) can be rewritten as
\begin{subequations}\label{eq:KM2}
	\begin{equation}\label{eq:KM2a}
	\dot{\zeta}_i=v_i+\omega_i+J_+S_+\sin{\left(\Phi_+-\zeta_i\right)}+J_-S_-\sin\left(\Phi_--\eta_i\right),
	\end{equation}
	\begin{equation}\label{eq:KM2b}
	\dot{\eta}_i=v_i-\omega_i+J_+S_-\sin{\left(\Phi_--\eta_i\right)}+J_-S_+\sin\left(\Phi_+-\zeta_i\right),
	\end{equation}
\end{subequations}
where $J_\pm=(J\pm K)/2$ are redefined coupling strengths and
\begin{equation}\label{eq:KM3}
	W_\pm=S_\pm e^{i\Phi_\pm}=\frac{1}{N}\sum_{j=1}^{N}e^{i(x_j\pm\theta_j)}.
\end{equation}
are order parameters characterizing the collective dynamics. If the phase and spatial variables are strongly correlated, i.e., $x_i=\pm\theta_i+c$ with $c$ a constant, then $S_\pm=1$. If $x_i$ and $\theta_i$ are fully uncorrelated, then $S_\pm=0$. In general, $S_\pm\in (0,1)$, and larger $S_\pm$ indicates stronger phase--space correlation.

From the perspective of multilayer complex networks, the dynamics can be interpreted as a two-layer network of coupled oscillators: Eq.~(\ref{eq:KM2a}) for the $\zeta$ layer and Eq.~(\ref{eq:KM2b}) for the $\eta$ layer. The two layers have the same size, and each node in one layer has a mirror node in the other. Unlike multilayer-network models in which mirror nodes are coupled adaptively through local (or global) order parameters~\cite{53,54,55,56,57,58}, our model employs a distinct interlayer coupling scheme. Specifically, each node in one layer (e.g., the $\zeta$ layer) receives signals from all nodes within the same layer via mean-field coupling of strength $J_+$, and also receives signals from the other layer through order-parameter-based coupling of strength $J_-$. As shown below, this layered-network viewpoint substantially simplifies the analysis of self-organized collective behaviors in swarmalators.

\begin{figure}[tbp]
\begin{center}
\includegraphics[width=0.85\linewidth]{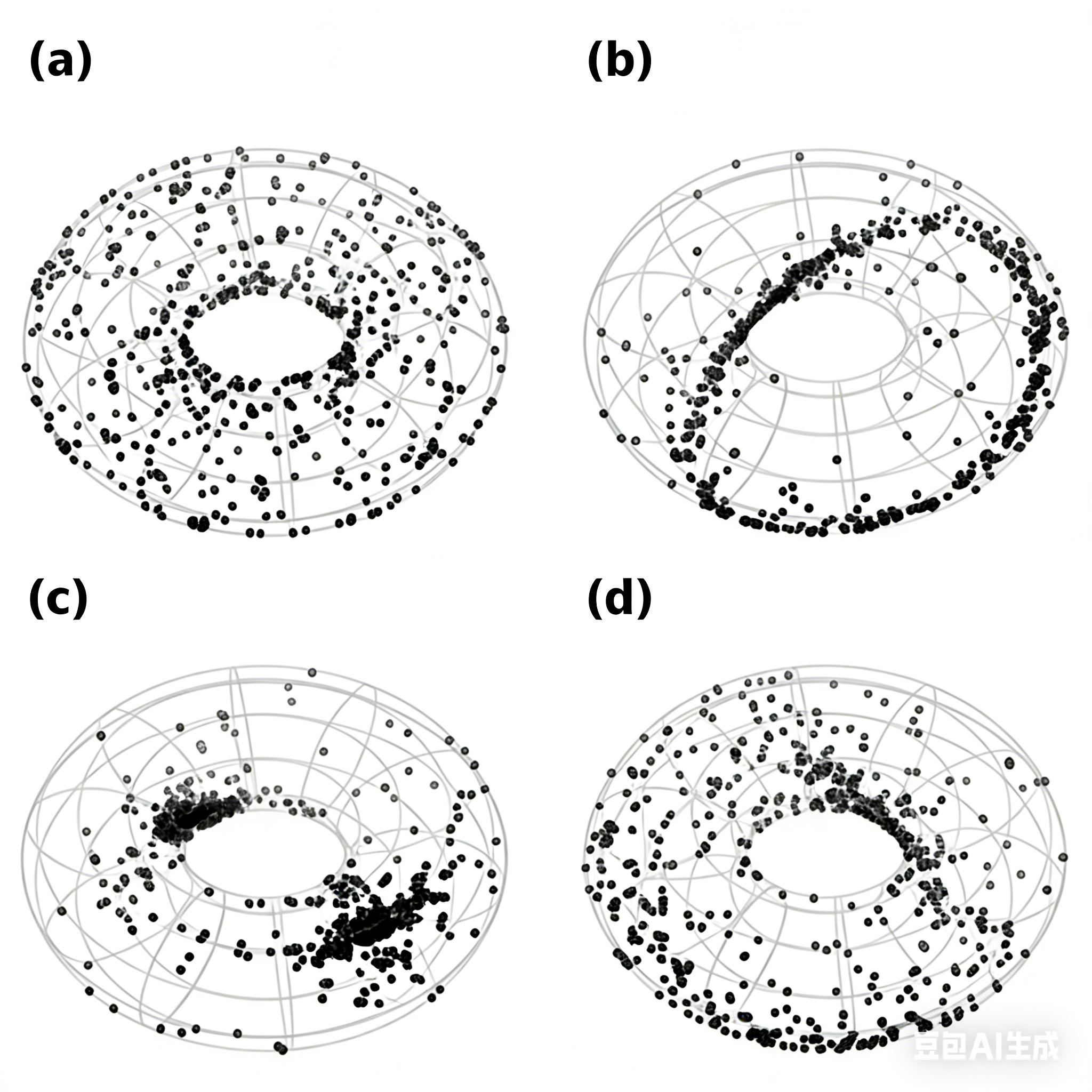}
\centering\caption{Schematic of the collective states observed in the one-dimensional swarmalator model.  (a) The desynchronization state.  (b) The phase-wave state.  (c) The synchronization state.  (d) The mixed state.  Please refer to Ref.~\cite{41} for a detailed description of the collective states.}
\vspace{-0.5cm}
\label{fig1}
\end{center}
\end{figure}

We now present the collective states observed in simulations. As reported in Ref.~\cite{41}, depending on the coupling strengths $J$ and $K$, four types of collective dynamics can occur: the desynchronization state, the phase-wave state, the synchronization state, and an intermediate mixed state. In the desynchronization state [Fig.~\ref{fig1}(a)], swarmalators are distributed nearly uniformly over the torus surface; coherent motion is absent in both space and phase, so $S_\pm\approx 0$. In the phase-wave state [Fig.~\ref{fig1}(b)], swarmalators align along a ring on the torus surface. Because $x_i\approx\mp\theta_i$ in this state, the order parameters satisfy $(S_+,S_-)=(S,0)$ or $(0,S)$ with $0<S<1$. According to Eqs.~(\ref{eq:KM2a}) and (\ref{eq:KM2b}), $(S_+,S_-)=(S,0)$ [or $(0,S)$] corresponds to partial synchronization in the $\zeta$ (or $\eta$) layer and desynchronization in the $\eta$ (or $\zeta$) layer. In the synchronization state [Fig.~\ref{fig1}(c)], swarmalators are partially synchronized in both layers, yielding $(S_+,S_-)=(S,S)$ with $0<S<1$. Finally, the mixed state [Fig.~\ref{fig1}(d)] consists of clustered synchronization on the torus surface and is typically observed in the transition region between the phase-wave and synchronization states. It is characterized by $(S_+,S_-)=(S_1,S_2)$ with $S_1\neq S_2$ and $S_{1,2}\in(0,1)$.

As discussed in Ref.~\cite{41}, the one-dimensional ring-structured model described by Eqs.~(\ref{eq:KM1a}) and (\ref{eq:KM1b}) can be viewed as a simplified version of the two-dimensional swarmalator model studied previously. It therefore serves as a basic setting for understanding collective dynamics in higher-dimensional swarmalator systems. Compared with high-dimensional counterparts, this simplified model enables efficient numerical simulations and permits more tractable theoretical analysis of the observed phenomena. Beyond its role as a reduced description, the one-dimensional model is also of independent interest, as it captures swarming behaviors observed in several real-world settings, such as the collective motion of vinegar eels and sperm cells, where edge-following tendencies constrain motion to ring-like orbits. These considerations motivate a detailed analysis of the one-dimensional model, including the nature of its collective states, their stability, the mechanisms underlying their formation, and their bifurcations in parameter space. Insights from this setting also provide theoretical guidance for nonequilibrium collective dynamics in active-matter systems.

We conclude this section by introducing the notations used throughout the paper. First, we consider the thermodynamic limit $N\rightarrow\infty$. Second, the natural frequencies $\omega_i$ and natural velocities $v_i$ are independent random variables drawn from unimodal symmetric distributions $g_\omega(\omega)$ and $g_v(v)$, respectively. Third, for any smooth bivariate function $f(v,\omega)$, we define the integral operator
\begin{equation}\label{eq:KM4}
\hat{g}f=\int_{-\infty}^{+\infty}dv\int_{-\infty}^{+\infty}d\omega g_v(v)g_\omega(\omega)f(v,\omega).
\end{equation}
Finally, throughout the paper, an overbar denotes complex conjugation and $^{\dagger}$ denotes the Hermitian conjugate, $\delta(x)$ denotes the Dirac function, $\Theta(x)$ denotes the Heaviside step function, and $\text{sgn}(x)$ is the sign function. For convenience, we also introduce two auxiliary parameters: $q=J_+ S$ and $p=J_- S$.

\section{Stability of the desynchronization state}\label{Sec III}

We first analyze the stability of the desynchronization state under general frequency--velocity distributions, with particular attention being paid to the structure of the eigenvalue spectrum and an explicit expression for the critical coupling strength. Because the swarmalator dynamics [Eqs.~(\ref{eq:KM1a}) and (\ref{eq:KM1b})] can be recast as a two-layer Kuramoto-type network [Eqs.~(\ref{eq:KM2a}) and (\ref{eq:KM2b})], we introduce a probability density function for each layer. For the $\zeta$ layer, we denote $\rho_+(v,\omega,\zeta,t)$ as the probability density of finding swarmalators with natural frequency $\omega$ and natural velocity $v$ in the interval $(\zeta,\zeta+d\zeta)$ at time $t$. The density is normalized as
\begin{equation}\label{eq:KM5}
\int_{0}^{2\pi} \rho_+(v,\omega,\zeta,t)d\zeta =1.
\end{equation}
Similarly, for the $\eta$ layer we introduce the normalized density $\rho_-(v,\omega,\eta,t)$.

In the thermodynamic limit, the deterministic layer dynamics implies the continuity equations
\begin{subequations}\label{eq:KM6}
	\begin{equation}\label{eq:KM6a}
		\frac{\partial \rho_+}{\partial t}+\frac{\partial (\rho_+ v_+)}{\partial \zeta} =0,
	\end{equation}
	\begin{equation}\label{eq:KM6b}
		\frac{\partial \rho_-}{\partial t}+\frac{\partial (\rho_- v_-)}{\partial \eta} =0,
	\end{equation}
\end{subequations}
with velocity fields
\begin{subequations}\label{eq:KM7}
	\begin{equation}\label{eq:KM7a}
		v_+=v+\omega +\frac{J_+}{2i}(W_+ e^{-i\zeta} - \overline{W}_+ e^{i\zeta})+\frac{J_-}{2i}(W_- e^{-i\eta}-\overline{W}_- e^{i\eta}),
	\end{equation}
	\begin{equation}\label{eq:KM7b}
		v_-=v-\omega +\frac{J_+}{2i}(W_- e^{-i\eta} - \overline{W}_- e^{i\eta})+\frac{J_-}{2i}(W_+ e^{-i\zeta}-\overline{W}_+ e^{i\zeta}) .
	\end{equation}
\end{subequations}
where we have omitted particle indices. In the continuum limit, the order parameters in Eq.~(\ref{eq:KM3}) become
\begin{subequations}\label{eq:KM8}
	\begin{equation}\label{eq:KM8a}
		W_+=\int_{0}^{2\pi} e^{i\zeta}\hat{g}\rho_+(v,\omega,\zeta,t)d\zeta ,
	\end{equation}
	\begin{equation}\label{eq:KM8b}
		W_-=\int_{0}^{2\pi} e^{i\eta}\hat{g}\rho_-(v,\omega,\eta,t)d\eta.
	\end{equation}
\end{subequations}

The desynchronization state corresponds to the uniform distribution
\begin{equation}\label{eq:KM9}
	\rho_+^{(0)}=\rho_-^{(0)}=\frac{1}{2\pi},
\end{equation}
i.e., $\zeta$ and $\eta$ are uniformly distributed over $[0,2\pi)$ and swarmalators are uniformly distributed over the torus. To assess its linear stability, we introduce perturbations
\begin{equation}\label{eq:KM10}
	\rho_\pm=\rho_\pm^{(0)}+\varepsilon \rho_\pm^{(1)},
\end{equation}
where $\varepsilon$ is a small parameter and $\rho_\pm^{(1)}$ are perturbation functions. Normalization implies the orthogonality conditions
\begin{subequations}\label{eq:KM11}
	\begin{equation}\label{eq:KM11a}
		\int_{0}^{2\pi}\rho_+^{(1)}(v,\omega,\zeta,t)d\zeta=0,
	\end{equation}
	\begin{equation}\label{eq:KM11b}
		\int_{0}^{2\pi}\rho_-^{(1)}(v,\omega,\eta,t)d\eta=0.
	\end{equation}
\end{subequations}
The corresponding order parameters satisfy
\begin{equation}\label{eq:KM12}
	W_\pm(t)=0+\varepsilon W_\pm^{(1)}(t),
\end{equation}
with
\begin{subequations}\label{eq:KM13}
	\begin{equation}\label{eq:KM13a}
		W_+^{(1)}=\int_{0}^{2\pi}e^{i\zeta}\hat{g}\rho_+^{(1)}(v,\omega,\zeta,t)d\zeta,
	\end{equation}
	\begin{equation}\label{eq:KM13b}
		W_-^{(1)}=\int_{0}^{2\pi}e^{i\eta}\hat{g}\rho_-^{(1)}(v,\omega,\eta,t)d\eta.
	\end{equation}
\end{subequations}
Substituting into Eqs.~(\ref{eq:KM6a}) and (\ref{eq:KM6b}) and retaining terms to first order in $\varepsilon$ yields
\begin{subequations}\label{eq:KM14}
	\begin{equation}\label{eq:KM14a}
		\dot{\rho}_+^{(1)}+\rho_+^{(0)}\frac{\partial v_+^{(1)}}{\partial \zeta}+v_+^{(0)}\frac{\partial \rho_+^{(1)}}{\partial \zeta}=0,
	\end{equation}
	\begin{equation}\label{eq:KM14b}
		\dot{\rho}_-^{(1)}+\rho_-^{(0)}\frac{\partial v_-^{(1)}}{\partial \eta}+v_-^{(0)}\frac{\partial \rho_-^{(1)}}{\partial \eta}=0,
	\end{equation}
\end{subequations}
where the unperturbed velocities are
\begin{equation}\label{eq:KM15}
	v_\pm^{(0)}=v\pm \omega,
\end{equation}
and the perturbed velocities are
\begin{subequations}\label{eq:KM16}
	\begin{equation}\label{eq:KM16a}
		v_+^{(1)}=\frac{J_+}{2i}(W_+^{(1)} e^{-i\zeta}-\overline{W}_+^{(1)} e^{i\zeta})+\frac{J_-}{2i}(W_-^{(1)} e^{-i\eta}-\overline{W}_-^{(1)} e^{i\eta}),
	\end{equation}
	\begin{equation}\label{eq:KM16b}
		v_-^{(1)}=\frac{J_+}{2i}(W_-^{(1)} e^{-i\eta}-\overline{W}_-^{(1)} e^{i\eta})+\frac{J_-}{2i}(W_+^{(1)} e^{-i\zeta}-\overline{W}_+^{(1)} e^{i\zeta}).
	\end{equation}
\end{subequations}
Consequently,
\begin{subequations}\label{eq:KM17}
	\begin{equation}\label{eq:KM17a}
		\frac{\partial v_+^{(1)}}{\partial \zeta}=-\frac{J_+}{2}(W_+^{(1)}e^{-i\zeta}+\overline{W}_+^{(1)}e^{i\zeta}),
	\end{equation}
	\begin{equation}\label{eq:KM17b}
		\frac{\partial v_-^{(1)}}{\partial \eta}=-\frac{J_+}{2}(W_-^{(1)}e^{-i\eta}+\overline{W}_-^{(1)}e^{i\eta}).
	\end{equation}
\end{subequations}

To solve the linearized system [Eqs.~(\ref{eq:KM14a}) and (\ref{eq:KM14b})], we expand the perturbations in Fourier series,
\begin{subequations}\label{eq:KM18}
	\begin{equation}\label{eq:KM18a}
		\rho_+^{(1)}=\frac{1}{2\pi}\sum_{n=-\infty}^{+\infty}\rho_{+,n}(v,\omega,t)e^{-in\zeta},
	\end{equation}
	\begin{equation}\label{eq:KM18b}
		\rho_-^{(1)}=\frac{1}{2\pi}\sum_{n=-\infty}^{+\infty}\rho_{-,n}(v,\omega,t)e^{-in\eta}.
	\end{equation}
\end{subequations}
The orthogonality conditions in Eqs.~(\ref{eq:KM11a}) and (\ref{eq:KM11b}) imply $\rho_{\pm,0}=0$, and reality implies $\rho_{\pm,-n}=\bar{\rho}_{\pm,n}$ for $n\neq 0$. The perturbed order parameters reduce to
\begin{equation}\label{eq:KM19}
	W_\pm^{(1)}(t)=\hat{g}\rho_{\pm,1}(v,\omega,t).
\end{equation}
Substituting into the continuity equations and projecting onto each Fourier mode yields
\begin{subequations}\label{eq:KM20}
	\begin{equation}\label{eq:KM20a}
		\dot{\rho}_{+,1}=iv_+^{(0)}\rho_{+,1} +\frac{J_+}{2}\hat{g}\rho_{+,1},
	\end{equation}
	\begin{equation}\label{eq:KM20b}
		\dot{\rho}_{-,1}=iv_-^{(0)}\rho_{-,1} +\frac{J_+}{2}\hat{g}\rho_{-,1},
	\end{equation}
	\begin{equation}\label{eq:KM20c}
		\dot{\rho}_{\pm,n}=inv_\pm^{(0)}\rho_{\pm,n}, \quad |n|>1.
	\end{equation}
\end{subequations}

For $|n|>1$, the eigenvalues are purely imaginary; these modes are neutrally stable and do not affect the stability of the desynchronization state. The stability is therefore determined by the $n=1$ mode, which can be written compactly as
\begin{equation}\label{eq:KM21}
	\frac{\mathrm{d}\mathbf{V}}{\mathrm{d}t}=\mathbf{\hat{J}}\mathbf{V},
\end{equation}
with $\mathbf{V}={(\rho_{+,1},\rho_{-,1})}^\top$ and
\begin{equation}\label{eq:KM22}
	\mathbf{\hat{J}}=
	\begin{pmatrix}
		iv_+^{(0)}+\frac{J_+}{2}\hat{g} & 0 \\
		0 & iv_-^{(0)}+\frac{J_+}{2}\hat{g}
	\end{pmatrix}.
\end{equation}

Let $\mathbf{V}(t)=\mathbf{V}_0e^{\lambda t}$, where $\mathbf{V}_0$ is determined by initial conditions, we hve
\begin{equation}\label{eq:KM23}
	\lambda \mathbf{V}_0=\mathbf{\hat{J}}\mathbf{V}_0.
\end{equation}
A nontrivial solution requires $|\lambda \mathbf{I}_2-\mathbf{\hat{J}}|=0$. Since $\mathbf{\hat{J}}$ is diagonal, the problem reduces to the scalar conditions. For smooth functions $c_{1,2}(v,\omega)$, we obtain
\begin{subequations}\label{eq:KM24}
	\begin{equation}\label{eq:KM24a}
		(\lambda - iv_+^{(0)}-\frac{J_+}{2}\hat{g})c_1(v,\omega)=0,
	\end{equation}
	\begin{equation}\label{eq:KM24b}
		(\lambda - iv_-^{(0)}-\frac{J_+}{2}\hat{g})c_2(v,\omega)=0,
	\end{equation}
\end{subequations}
and hence
\begin{subequations}\label{eq:KM25}
	\begin{equation}\label{eq:KM25a}
		c_1=\frac{\frac{J_+}{2}\hat{g}c_1}{\lambda-iv_+^{(0)}},
	\end{equation}
	\begin{equation}\label{eq:KM25b}
		c_2=\frac{\frac{J_+}{2}\hat{g}c_2}{\lambda-iv_-^{(0)}}.
	\end{equation}
\end{subequations}
Applying $\hat{g}$ to both sides gives the self-consistency equations
\begin{subequations}\label{eq:KM26}
	\begin{equation}\label{eq:KM26a}
		1=\frac{J_+}{2}\hat{g}(\lambda-iv_+^{(0)})^{-1},
	\end{equation}
	\begin{equation}\label{eq:KM26b}
		1=\frac{J_+}{2}\hat{g}(\lambda-iv_-^{(0)})^{-1}.
	\end{equation}
\end{subequations}
The continuous spectrum of the desynchronization state therefore is
\begin{equation}\label{eq:KM27}
	\sigma_c^{\text{desyn}} = \{\lambda = i v_+^{(0)} \} \cup \{ \lambda = i v_-^{(0)} \},
\end{equation}
which lies on the imaginary axis.

We next study Eqs.~(\ref{eq:KM26a}) and (\ref{eq:KM26b}) for $\lambda\notin\sigma_c^{\text{desyn}}$. We start by analyzing Eq.~(\ref{eq:KM26a}). By the definition of $\hat{g}$, we have
\begin{equation}\label{eq:KM28}
	\frac{2}{J_+}=\int_{-\infty}^{+\infty} \int_{-\infty}^{+\infty} \frac{g_v(v)g_\omega(\omega)}{\lambda-i(v+\omega)}dv d\omega.
\end{equation}
Letting $x=v+\omega$, this becomes
\begin{equation}\label{eq:KM29}
	\frac{2}{J_+}=\int_{-\infty}^{+\infty}\frac{\tilde{g}_+(x)}{\lambda -ix}dx,
\end{equation}
where
\begin{equation}\label{eq:KM30}
	\tilde{g}_+(x)=\int_{-\infty}^{+\infty} g_v(x-\omega)g_\omega(\omega)d\omega
\end{equation}
is the convolution distribution. Similarly, Eq.~(\ref{eq:KM26b}) yields
\begin{equation}\label{eq:KM31}
	\frac{2}{J_+}=\int_{-\infty}^{+\infty}\frac{\tilde{g}_-(x)}{\lambda -ix}dx,
\end{equation}
with
\begin{equation}\label{eq:KM32}
	\tilde{g}_-(x)=\int_{-\infty}^{+\infty} g_v(v)g_\omega(x-v)dv.
\end{equation}
Because $g_v(v)$ and $g_\omega(\omega)$ are even, we have
\begin{equation}\label{eq:KM33}
	\tilde{g}_+(x)=\tilde{g}_-(x)=\tilde{g}(x),
\end{equation}
where $\tilde{g}(x)$ is also even. Hence both self-consistency conditions reduce to
\begin{equation}\label{eq:KM34}
	\frac{2}{J_+}=\int_{-\infty}^{+\infty}\frac{\tilde{g}(x)}{\lambda -ix}dx.
\end{equation}
If $\tilde{g}(x)$ is unimodal and symmetric, then $\lambda$ is real (as in the classical Kuramoto model). Using the identity $\lim_{\lambda \rightarrow 0^+}(\lambda-ix)^{-1}=\pi \delta(x)$, we obtain the critical coupling strength beyond which the desynchronization is unstable
\begin{equation}\label{eq:KM35}
    J_{+,c}=\frac{2}{\pi \tilde{g}(0)}.
\end{equation}

We find that $J_-$ has no effect on the stability of the desynchronization state. At first glance, one might expect that the coupling strengths $J_+$ and $J_-$ should play identical functional roles, since they correspond to the weights of the order parameters for the two-layer coupled oscillator system, and should at least exhibit the same influence on the critical threshold of desynchronization instability. Nevertheless, this is not true. Mathematically, the term $J_-$ in fact corresponds to a nonlinear perturbation, and therefore does not contribute to the linear stability analysis. Physically, this counterintuitive phenomenon can be interpreted as follows: $J_-$ essentially describes the driving coupling strength from one layer of oscillators to the other. Specifically, taking the $\zeta$ layer as an example, the term associated with $J_-$ corresponds exactly to the external driving of the variable $\zeta$ imposed by the input variable $\eta$. When the system resides in the desynchronization state, the oscillators within each layer oscillate freely at their natural frequencies, i.e., $\eta \sim (\omega - v)t$. In this case, the external driving acting on $\zeta$ takes the form of a high-frequency oscillation $J_- \sin(\omega -v)t$. In the long-time limit, such an oscillatory term vanishes under temporal averaging. This elucidates why $J_-$ exerts no substantial influence on the stability threshold of the asynchronous state. However, as will be shown in the following sections, $J_-$ plays a critical role in determining the existence and stability intervals of the other three collective states.

In summary, we have analyzed the stability of the desynchronization state by first mapping the swarmalator model to a two-layer Kuramoto network, then introducing layer-wise probability densities, and finally deriving self-consistent equations for the eigenvalues of the associated linear operator. The continuous spectrum consists of two sets of purely imaginary eigenvalues, associated with the effective natural frequencies $v+\omega$ and $v-\omega$ in the $\zeta$ and $\eta$ layers, respectively. The discrete spectrum is doubly degenerate due to the mirror symmetry of the two-layer structure. The main result is given by Eq.~(\ref{eq:KM35}), which defines the critical coupling strength beyond which the desynchronization state becomes unstable. Unlike earlier special cases, Eq.~(\ref{eq:KM35}) applies to a general convolutional frequency--velocity distribution $\tilde{g}(x)$, making it suitable for exploring the interplay between spatial and phase dynamics in swarmalators.

Equation~(\ref{eq:KM35}) can be interpreted as follows. In the desynchronization state, mirror symmetry implies $W_\pm=0$, so the two layers decouple in Eqs.~(\ref{eq:KM2a}) and (\ref{eq:KM2b}). Each layer then follows Kuramoto-type mean-field dynamics with effective natural frequency $v+\omega$ for the upper layer and $v-\omega$ for the lower layer, so the stability reduces to that of the incoherent state in the classical Kuramoto model. Moreover, for independent random variables, the distributions of the sum and difference are given by convolutions of the original distributions; if both variables are symmetric, then the sum and difference share the same convolution distribution. This explains why Eq.~(\ref{eq:KM35}) generalizes the classical Kuramoto threshold.

We next verify Eq.~(\ref{eq:KM35}) with several examples. For Cauchy--Lorentz distributions,
\begin{equation}\label{eq:KM36}
	g_v(v)=\frac{\Delta_v}{\pi}\frac{1}{v^2 +\Delta_v^2},  g_\omega(\omega)=\frac{\Delta_\omega}{\pi}\frac{1}{\omega^2 +\Delta_\omega^2},
\end{equation}
where $\Delta_v$ and $\Delta_\omega$ are half-widths, direct calculation gives
\begin{equation}\label{eq:KM37}
	\tilde{g}_{LL}(x)=\frac{\Delta_v +\Delta_\omega}{\pi}\frac{1}{x^2 +(\Delta_v+\Delta_\omega)^2}.
\end{equation}
Substituting Eq.~(\ref{eq:KM37}) into Eq.~(\ref{eq:KM35}) yields
\begin{equation}\label{eq:KM38}
J_{+,c}=2(\Delta_v+\Delta_\omega).
\end{equation}

For Gaussian distributions,
\begin{equation}\label{eq:KM39}
	g_v(v)=\frac{1}{\sqrt{2\pi \sigma_v^2}}e^{-\frac{v^2}{2\sigma_v^2}},  g_\omega(\omega)=\frac{1}{\sqrt{2\pi \sigma_\omega^2}}e^{-\frac{\omega^2}{2\sigma_\omega^2}},
\end{equation}
where $\sigma_v^2$ and $\sigma_\omega^2$ are variances, the convolution distribution is
\begin{equation}\label{eq:KM40}
	\tilde{g}_{GG}(x)=\frac{1}{\sqrt{2\pi (\sigma_v^2 + \sigma_\omega^2)}} e^{-\frac{x^2}{2(\sigma_v^2 + \sigma_\omega^2)}}.
\end{equation}
Substituting into Eq.~(\ref{eq:KM35}) gives
\begin{equation}\label{eq:KM41}
	J_{+,c} =\frac{2\sqrt{2(\sigma_v^2 +\sigma_\omega^2)}}{\sqrt{\pi}}.
\end{equation}

The two examples above involve stable distributions, in the sense that linear combinations of independent random variables preserve the distributional family up to parameter changes. For instance, for zero-mean Cauchy--Lorentz distributions, the sum and difference remain zero-mean Cauchy--Lorentz with half-width $\Delta_v+\Delta_\omega$, while for zero-mean Gaussians the sum and difference remain Gaussian with variance $\sigma_v^2+\sigma_\omega^2$.

\begin{figure}[tbp]
\begin{center}
\includegraphics[width=1\linewidth]{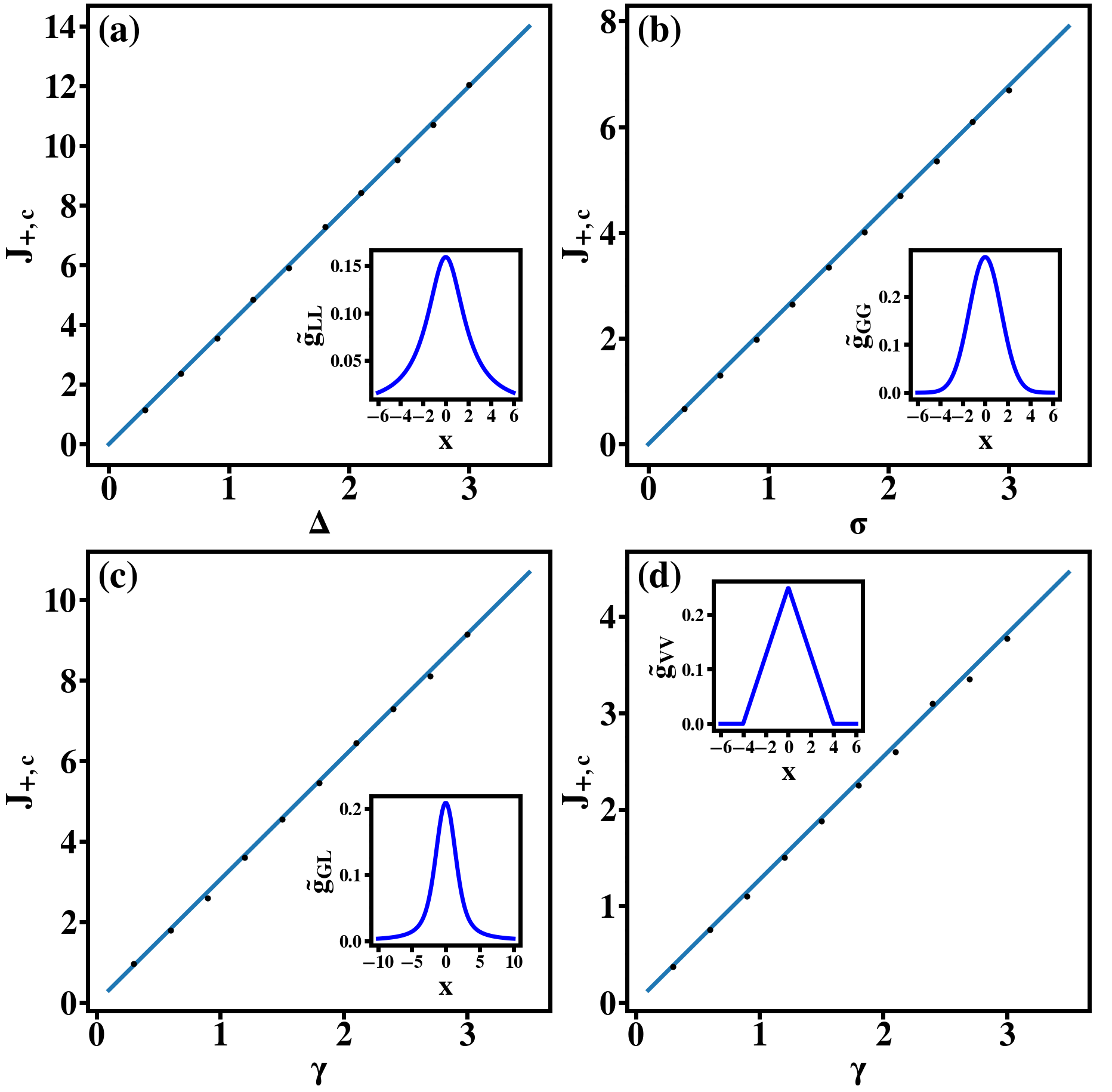}
\caption{The variation of the critical coupling strength, $J_+$, for the desynchronization state with respect to the frequency--velocity distribution parameters. (a) Lorentzian--Lorentzian distribution combination with parameters $\Delta_v =\Delta_\omega=\Delta$. (b) Gaussian--Gaussian distribution combination with parameters $\sigma_v =\sigma_\omega=\sigma$. (c) Gaussian--Lorentzian distribution combination with parameters $\sigma_v=\Delta_\omega=\gamma$. (d) Uniform--uniform distribution combination. Insets are schematic of the convolution distributions. Blue curves are the theoretical results predicted by Eq.~ (\ref{eq:KM35}). Black dots are the results obtained by numerical simulations. In simulations, the system size is chosen as $N=5\times 10^4$.}
\vspace{-0.2cm}
\label{fig2}
\end{center}
\end{figure}

We next apply Eq.~(\ref{eq:KM35}) to a more involved case, namely a Gaussian--Lorentz combination, for which the convolution yields the Voigt distribution and has no closed analytical form. For uniform distributions,
\begin{equation}\label{eq:KM43}
	g_v(v)=\frac{1}{2\gamma} \Theta(\gamma-|v|), 	g_\omega(\omega)=\frac{1}{2\gamma} \Theta(\gamma-|\omega|),
\end{equation}
we obtain
\begin{equation}\label{eq:KM42}
	\tilde{g}_{GL}(0)=\frac{1}{\sqrt{2\pi \sigma_v^2}}e^{\frac{\Delta_\omega^2}{2\sigma_v^2}}\text{erfc}(\frac{\Delta_\omega}{\sqrt{2}\sigma_v}),
\end{equation}
where $\Theta(x)$ is the Heaviside step function and $\gamma$ is the width of the uniform distributions. In this case, the distribution is not stable, because the sum (or difference) of two uniform distributions is not uniform. Indeed, direct calculation shows that the convolution is triangular,
\begin{equation}\label{eq:KM44}
	\tilde{g}_{UU}(x)=\frac{2\gamma-|x|}{4\gamma^2} \Theta(2\gamma-|x|).
\end{equation}

Finally, substituting Eqs.~(\ref{eq:KM42}) and (\ref{eq:KM44}) into Eq.~(\ref{eq:KM35}) yields the critical coupling strength for the Lorentzian--Lorentzian, Gaussian--Gaussian, Gaussian--Lorentzian, and uniform--uniform combinations. Figure~\ref{fig2} compares these theoretical predictions with numerical simulations, showing excellent agreement.

\section{Connection with the high-dimensional agent model}\label{Sec IV}

A central question in the one-dimensional ring-structured swarmalator model is why the high-dimensional microscopic dynamics described by Eqs.~(\ref{eq:KM1a}) and (\ref{eq:KM1b}) exhibits a universal low-dimensional macroscopic invariant manifold. Addressing this question not only clarifies the intrinsic relation between swarmalator models and high-dimensional agent models, but also places these studies within the general framework of the Ott--Antonsen ansatz. Building on the results in Sec.~\ref{Sec III}, we now analyze the connection between the one-dimensional swarmalator model and Tanaka's high-dimensional agent model.

As noted above, the one-dimensional swarmalator model can be interpreted as motion on the surface of a two-dimensional torus. This observation naturally links it to high-dimensional agent models. Introducing the complex coordinates $z_+=e^{i\zeta}$ and $z_-=e^{i\eta}$, we map Eqs.~(\ref{eq:KM2a}) and (\ref{eq:KM2b}) from the torus $(\zeta,\eta)\in\mathbb{R}^2$ to a complex space, obtaining
\begin{subequations}\label{eq:KM45}
	\begin{equation}\label{eq:KM45a}
		\begin{aligned}
		\dot{z}_+=&i(v+\omega)z_+ +\frac{J_+}{2}(W_+ -z_+ \overline{W}_+ z_+) \\
		&+\frac{J_-}{2}(z_+ W_- \overline{z}_- -z_+\overline{W}_- z_-),
		\end{aligned}
	\end{equation}
	\begin{equation}\label{eq:KM45b}
		\begin{aligned}
		\dot{z}_-=&i(v-\omega)z_- +\frac{J_+}{2}(W_- -z_- \overline{W}_- z_-) \\
		&+\frac{J_-}{2}(z_- W_+ \overline{z}_+ -z_-\overline{W}_+ z_+).	
		\end{aligned}
	\end{equation}
\end{subequations}
Thus, motion on the torus is represented as motion on a two-dimensional complex sphere. Defining the complex vector $\mathbf{X}=(z_+,z_-)^\top\in\mathbb{C}^2$, Eq.~(\ref{eq:KM45}) can be written compactly as
\begin{equation}\label{eq:KM46}
	\dot{\mathbf{X}}=\mathbf{\Omega}\mathbf{X}+\mathbf{G}-\mathbf{X}(\mathbf{G}^\dagger\mathbf{X}),
\end{equation}
where the complex mean-field vector is $\mathbf{G}=\frac{J_+}{2}(W_+,W_-)^\top$ and $\mathbf{\Omega}=\text{diag}(\Omega_{11},\Omega_{22})$, with
\begin{subequations}\label{eq:KM47}
	\begin{equation}\label{eq:KM47a}
		\Omega_{11}=i(v+\omega)+\frac{1}{2}(J_+ -J_-)\overline{W}_-z_-+\frac{J_-}{2}W_-\overline{z}_- ,
	\end{equation}
	\begin{equation}\label{eq:KM47b}
		\Omega_{22}=i(v-\omega)+\frac{1}{2}(J_+ -J_-)\overline{W}_+z_++\frac{J_-}{2}W_+\overline{z}_+ .
	\end{equation}
\end{subequations}
Remarkably, Eq.~(\ref{eq:KM46}) has the same structure as the generalized agent model proposed by Tanaka (see Refs.~\cite{52,59} for details). In his pioneering work, Tanaka extended coupled phase oscillators (entities on the unit circle in $\mathbb{C}$) to agents moving on a high-dimensional complex sphere $\mathbb{C}^m$. Since each complex coordinate corresponds to a two-dimensional real vector, subsequent studies often describe the same dynamics as motion on a high-dimensional real hypersphere~\cite{60,61,62,63,64,65,66,67,68,69}, commonly referred to as the high-dimensional Kuramoto model.

In Tanaka's model, the rotation matrix $\mathbf{\Omega}$ is required to be anti-Hermitian, $\mathbf{\Omega}^{\dagger}=-\mathbf{\Omega}$, to preserve the modulus of the complex vector. For example, the classical Kuramoto model corresponds to $m=1$ with $\mathbf{\Omega}=i\omega$. In contrast, for Eq.~(\ref{eq:KM46}) the evolution of each component [Eqs.~(\ref{eq:KM45a}) and (\ref{eq:KM45b})] preserves $|z_\pm|=1$ automatically, and $\mathbf{\Omega}$ is diagonal; hence no anti-Hermitian constraint is required. In this sense, the one-dimensional ring-structured swarmalator model can be viewed as a generalization of Tanaka's framework.

According to Tanaka's theory~\cite{52}, for vector dynamics of the form in Eq.~(\ref{eq:KM46}), identical agents (i.e., the same rotation matrix for all agents) evolve along a high-dimensional Möbius transformation (group orbit). In particular, if initial conditions are uniformly distributed on the unit complex sphere, the macroscopic order-parameter vector $\mathbf{P}=N^{-1}\sum_{j=1}^{N}\mathbf{X}_j$ obeys the same equation as a single agent, obtained by replacing $\mathbf{X}$ with $\mathbf{P}$ in Eq.~(\ref{eq:KM46}). This result generalizes the classical Kuramoto model for identical oscillators~\cite{70,71,72}. For heterogeneous agents, the rotation matrix becomes index-dependent with a prescribed distribution. As in the classical Kuramoto model, one may interpret $\mathbf{P}$ as a local order parameter formed by agents with the same rotation matrix; this local order parameter satisfies the same form of equation, leading to a high-dimensional generalization of the Ott--Antonsen reduction~\cite{73,74}. Consequently, the conceptual route from Möbius transformations to the Ott--Antonsen manifold extends naturally to high-dimensional heterogeneous agents. Motivated by this viewpoint, we now derive the low-dimensional invariant manifold for the swarmalator model in a layered-network formulation, starting from the continuity equation.

In Fourier space, the density in each layer can be expanded as
\begin{subequations}\label{eq:KM48}
	\begin{equation}\label{eq:KM48a}
		\rho_+=\frac{1}{2\pi}\sum_{n=-\infty}^{+\infty}\alpha_n(v,\omega,t)e^{-in\zeta},
	\end{equation}
	\begin{equation}\label{eq:KM48b}
		\rho_-=\frac{1}{2\pi}\sum_{n=-\infty}^{+\infty}\beta_n(v,\omega,t)e^{-in\eta}.
	\end{equation}
\end{subequations}
Normalization implies $\alpha_0=\beta_0=1$. Since $\rho_\pm$ are real, the Fourier coefficients satisfy $\alpha_{-n}=\overline{\alpha}_n$ and $\beta_{-n}=\overline{\beta}_n$. We further define
\begin{subequations}\label{eq:KM49}
	\begin{equation}\label{eq:KM49a}
		\Delta_+=\frac{J_-}{2i}(W_-e^{-i\eta}-\overline{W}_- e^{i\eta}),
	\end{equation}
	\begin{equation}\label{eq:KM49b}
		\Delta_-=\frac{J_-}{2i}(W_+ e^{-i\zeta}-\overline{W}_+ e^{i\zeta}).
	\end{equation}
\end{subequations}
Substituting Eqs.~(\ref{eq:KM48a}) and (\ref{eq:KM49a}) into the continuity equation (\ref{eq:KM6a}) and projecting onto Fourier modes yields
\begin{equation}\label{eq:KM50}
	\scalebox{0.8}{$\dot{\alpha}_n-in(v+\omega)\alpha_n-\frac{nJ_+}{2}W_+ \alpha_{n-1}+\frac{nJ_+}{2}\overline{W}_+\alpha_{n+1}-in\Delta_+ \alpha_n=0,$}
\end{equation}
showing a cascade coupling across Fourier modes. The modes can be decoupled via the Ott--Antonsen ansatz
\begin{equation}\label{eq:KM51}
	\alpha_n(v,\omega,t) = \alpha(v,\omega,t)^n,
\end{equation}
which corresponds to a Poisson-kernel density. Substituting Eq.~(\ref{eq:KM51}) into Eq.~(\ref{eq:KM50}) yields the evolution equation for the first Fourier mode,
\begin{equation}\label{eq:KM52}
	\dot{\alpha}=i(v+\omega)\alpha+\frac{J_+}{2}(W_+ -\overline{W}_+ \alpha^2)+i\Delta_+ \alpha.
\end{equation}
Similarly, for the $\eta$ layer we obtain
\begin{equation}\label{eq:KM53}
	\dot{\beta}=i(v-\omega)\beta+\frac{J_+}{2}(W_- -\overline{W}_- \beta^2)+i\Delta_- \beta.
\end{equation}

To eliminate $\Delta_\pm$, we apply the operators $\langle\cdot\rangle_\mp=\int_{0}^{2\pi}(\cdot)\rho_\mp\, d\eta(d\zeta)$ to Eqs.~(\ref{eq:KM52}) and (\ref{eq:KM53}). Using normalization, we arrive at the closed system
\begin{subequations}\label{eq:KM54}
	\begin{equation}\label{eq:KM54a}
	\dot{\alpha}=i(v+\omega)\alpha+\frac{J_+}{2}(W_+ -\overline{W}_+ \alpha^2)+\frac{J_-}{2}(W_- \overline{\beta}-\overline{W}_- \beta) \alpha,
	\end{equation}
	\begin{equation}\label{eq:KM54b}
	\dot{\beta}=i(v-\omega)\beta+\frac{J_+}{2}(W_- -\overline{W}_- \beta^2)+\frac{J_-}{2}(W_+ \overline{\alpha}-\overline{W}_+ \alpha) \beta,
	\end{equation}
\end{subequations}
together with the closure relations
\begin{subequations}\label{eq:KM55}
	\begin{equation}\label{eq:KM55a}
		W_+(t)=\hat{g}\alpha(v,\omega,t),
	\end{equation}
	\begin{equation}\label{eq:KM55b}
		W_-(t)=\hat{g}\beta(v,\omega,t).
	\end{equation}
\end{subequations}
Equations~(\ref{eq:KM54})--(\ref{eq:KM55}) define the Ott--Antonsen invariant manifold for the swarmalator model.

Taken together, starting from the continuity equation, we have derived an explicit low-dimensional invariant manifold for the swarmalator model. Interpreting the first Fourier coefficients $\alpha(v,\omega,t)$ and $\beta(v,\omega,t)$ as local order parameters formed by swarmalators with given $(v,\omega)$, and defining $\mathbf{P}=(\alpha,\beta)^\top$, Tanaka's theory implies that $\mathbf{P}$ satisfies Eq.~(\ref{eq:KM46}), which is precisely equivalent to Eqs.~(\ref{eq:KM54a}) and (\ref{eq:KM54b}). We therefore conclude that Tanaka's agent model provides a bridge between high-dimensional agent dynamics and the reduced swarmalator description. The associated generalized Möbius transformation (and the generalized Ott--Antonsen manifold) provides a theoretical foundation for dimension reduction in such systems, offering a useful perspective for analyzing spatiotemporal pattern formation in active matter.

It is worth noting that we have only provided a formal justification for the validity of the Ott-Antonsen (OA) ansatz. Fundamentally, whether in classical Kuramoto-like coupled phase oscillator systems (scalar) or in the recently highlighted high-dimensional agent systems (vector), the mathematical foundation enabling the dimensional reduction of these systems is rooted in their hidden symmetries—specifically, the existence of numerous constants of motion. Research has established that such coupled-oscillator systems can generally be mapped to Riccati equations, in which the dynamical evolution is governed by a family of Möbius transformations (fractional linear mappings). Specifically, when the constants of motion follow a uniform measure, the Möbius transformation naturally reduces to the OA ansatz. In other words, the OA ansatz represents merely one specific class of invariant manifolds within the system, rather than the only one. This inherently prompts a discussion regarding the attractiveness, and even the basin of attraction, of the OA manifold. Particularly, when the natural frequencies of the oscillators are identical, the system exhibits quasi-Hamiltonian characteristics (partial integrability), under which the OA manifold ceases to be globally attractive~\cite{72}. A quantitative analysis of the OA manifold's attractiveness falls outside the scope of the present study. For further discussions on this specific topic, we direct interested readers to Ref. ~\cite{95,96}. Nevertheless, we firmly believe that exploring the attractiveness of the OA manifold within swarmalator systems, and subsequently constructing a theoretical framework that transcends the OA ansatz to unveil the emergence mechanisms of broader spatiotemporal collective patterns, stands as a highly challenging yet profoundly significant avenue for future research.

\section{Applications of the low-dimensional manifold}\label{Sec V}

We next use the Ott--Antonsen manifold obtained in Sec.~\ref{Sec IV} [Eq.~(\ref{eq:KM54})] to address several open issues raised in previous studies of swarmalators.

\subsection{The desynchronization state}

A direct application of the Ott--Antonsen manifold is the stability analysis of the desynchronization state. In Ref.~\cite{41}, based on the macroscopic observation $W_\pm=0$, it was argued that $\alpha=e^{i(v+\omega)t}$ and $\beta=e^{i(v-\omega)t}$, and this assumption was used as the basis for stability analysis; similar assumptions were adopted in Refs.~\cite{50,51}. Physically, time-dependent factors can indeed be removed by long-time averaging. Mathematically, however, the desynchronization state is a steady state corresponding to $\rho_\pm=1/(2\pi)$, and under the Ott--Antonsen ansatz [Eq.~(\ref{eq:KM51})] all Fourier coefficients are time independent. In fact, the desynchronization state corresponds to $\alpha=\beta=0$. While $W_\pm=0$ is valid, it is also simply the trivial fixed point of Eq.~(\ref{eq:KM54}). Therefore, the stability of the desynchronization state reduces to the linear stability of the origin for Eq.~(\ref{eq:KM54}).

Accordingly, we set $\alpha=0+\delta\alpha$ and $\beta=0+\delta\beta$, and write $\delta W_+=\hat{g}\delta\alpha$ and $\delta W_-=\hat{g}\delta\beta$. Substituting into Eq.~(\ref{eq:KM54}) and retaining only linear terms yields
\begin{subequations}\label{eq:KM56}
	\begin{equation}\label{eq:KM56a}
		\delta\dot{\alpha}=i(v+\omega)\delta\alpha +\frac{J_+}{2}\delta W_+,
	\end{equation}
	\begin{equation}\label{eq:KM56b}
		\delta\dot{\beta}=i(v-\omega)\delta\beta +\frac{J_+}{2}\delta W_-.
	\end{equation}
\end{subequations}
Following the same procedure as in Sec.~\ref{Sec III}, one recovers the self-consistency relations for the eigenvalues [Eqs.~(\ref{eq:KM26a}) and (\ref{eq:KM26b})], and the critical coupling strength follows immediately.

Note that the instability threshold depends only on $J_+$, which denotes the intra-layer coupling strength. This is because the inter-layer terms proportional to $J_-$ contribute at higher order near $\alpha=\beta=0$ and can be neglected in the linear stability analysis. In particular, whether one starts from the continuity equation or from the Ott--Antonsen manifold, the critical coupling strength [Eq.~(\ref{eq:KM35})] is independent of the specific forms of $g_\omega(\omega)$ and $g_v(v)$; this point will be emphasized repeatedly below.

\subsection{The Phase-wave state}

We next analyze the phase-wave state shown in Fig.~\ref{fig1}(b), characterized by order parameters $(S,0)$ or $(0,S)$. This state is highly nontrivial: oscillators in one layer remain disordered, while oscillators in the other layer are partially synchronized. As an intermediate state between desynchronization and synchronization, it plays a key role in understanding the transitions and bifurcations of the system dynamics in the parameter space. Phenomenologically, this partially coherent state resembles generalized chimera states in coupled oscillators~\cite{75}, in which a subset of oscillators synchronizes while the remainder stays incoherent. Owing to mirror symmetry, the phase-wave state is bistable: $(S,0)$ and $(0,S)$ occur with equal probability for random initial conditions. In this subsection, we analyze the existence, stability, and bifurcation structure of this state for general frequency--velocity distributions.

\subsubsection{Existence of the phase-wave state}

Concerning symmetry, we focus on only the $(S,0)$ branch, for which $W_+=S$ and $W_-=0$. On the Ott--Antonsen manifold, this corresponds to $\alpha(v,\omega,t)=\alpha_{pw}(v,\omega)$ and $\beta=0$ (in contrast to the time-dependent expression stated in Ref.~\cite{41}). The steady-state condition becomes
\begin{equation}\label{eq:KM57}
	i(v+\omega)\alpha_{pw} + \frac{q}{2}(1 - \alpha_{pw}^2) = 0.
\end{equation}
Solving this quadratic equation gives
\begin{equation}\label{eq:KM58}
	\alpha= \pm\sqrt{1 - (\frac{v+\omega}{q})^2}+ i \frac{v+\omega}{q}.
\end{equation}
Depending on $1-(v+\omega)^2/q^2$, we obtain
\begin{equation}\label{eq:KM59}
	\alpha_{pw} = 
	\begin{cases}
	\scalebox{0.9}{$\sqrt{1 - \left(\dfrac{v+\omega}{q}\right)^2} + i \dfrac{v+\omega}{q}, \quad |v+\omega| \leq q,$} \\
	\scalebox{0.9}{$-i\operatorname{sgn}(v+\omega)  \sqrt{\left(\dfrac{v+\omega}{q}\right)^2 - 1} + i \dfrac{v+\omega}{q}, \quad |v+\omega| > q.$}		
	\end{cases}
\end{equation}
The first case corresponds to locked oscillators, for which $|\alpha_{pw}(v,\omega)|=1$; the positive real part selects the stable branch. The second case corresponds to drifting oscillators and ensures $|\alpha_{pw}|\le 1$. As we will show below, both locked and drifting contributions are relevant for the stability of the phase-wave state.

Using the definition of $W_+$ in Eq.~(\ref{eq:KM55a}), we obtain the self-consistency condition
\begin{equation}\label{eq:KM60}
	\frac{1}{J_+}=\int_{-1}^{1}\sqrt{1-x^2}\tilde{g}(qx)dx.
\end{equation}
By symmetry, neither the drifting population nor the imaginary part of the locked population contributes to the real order parameter.

Equation~(\ref{eq:KM60}) generalizes the Kuramoto self-consistency relation by replacing the frequency distribution with the convolution distribution $\tilde{g}(x)$. This is also consistent with the layered-network viewpoint: in the phase-wave state, the two layers are effectively decoupled, with the $\zeta$ layer following Kuramoto-type dynamics with effective frequency $v+\omega$ distributed according to $\tilde{g}(x)$, while oscillators in the $\eta$ layer drift with effective frequency $v-\omega$ and remain incoherent. Consequently, $J_-$ does not appear in Eq.~(\ref{eq:KM60}). As in the classical Kuramoto model, the onset of the phase-wave state corresponds to $q\rightarrow 0^+$, yielding the critical coupling strength $J_{+,c}=2/(\pi\tilde{g}(0))$, identical to the threshold for destabilizing the desynchronization state [Eq.~(\ref{eq:KM35})]. For the Lorentz--Lorentz case, one obtains explicitly
\begin{equation}\label{eq:KM61}
	S=\sqrt{1-\frac{2(\Delta_v+\Delta_\omega)}{J_+}}.
\end{equation}
This expression is in complete agreement with the result reported in Ref. ~\cite{41}, where it has been well validated numerically. For other distribution combinations, closed forms are generally unavailable, but standard Kuramoto arguments still apply to universal behaviors such as critical scaling~\cite{76,77,78} and susceptibility~\cite{79,80,81,82}.

Thus, both the onset of the phase-wave state and the self-consistency equation~(\ref{eq:KM60}) depend only on the intra-layer coupling $J_+$, and are independent of the inter-layer coupling $J_-$. In particular, when $J_+>J_{+,c}$ the desynchronization state loses stability, and the system organizes into a phase-wave state with order parameter $S$ determined solely by $J_+$. As $J_\pm$ increases further, the incoherent $\eta$ layer may also become unstable, which then feeds back onto the $\zeta$ layer and ultimately destabilizes the phase-wave state.

\subsubsection{Stability of the phase-wave state} 

We now analyze the stability of the phase-wave state. Introducing perturbations around $(\alpha_{pw},0)$,
\begin{subequations}\label{eq:KM62}
	\begin{equation}\label{eq:KM62a}
	\alpha(v,\omega,t)=\alpha_{pw}+\varepsilon a(v,\omega,t),
	\end{equation}
	\begin{equation}\label{eq:KM62b}
		\beta(v,\omega,t)=0+\varepsilon b(v,\omega,t),
	\end{equation}
\end{subequations}
the perturbed order parameters [from Eq.~(\ref{eq:KM55})] are
\begin{subequations}\label{eq:KM63}
	\begin{equation}\label{eq:KM63a}
		W_+=S+\varepsilon \hat{g}a, \qquad \overline{W}_+=S+\varepsilon \hat{g}\overline{a},
	\end{equation}
	\begin{equation}\label{eq:KM63b}
		W_-=\varepsilon \hat{g}b, \qquad \overline{W}_-=\varepsilon \hat{g}\overline{b}.
	\end{equation}
\end{subequations}
Defining $\mathbf{V}=(a,\overline{a},b)^\top\in\mathbb{C}^3$ and substituting into Eq.~(\ref{eq:KM54}), we obtain the linearized system (to first order in $\varepsilon$),
\begin{equation}\label{eq:KM64}
	\frac{\mathrm{d}\mathbf{V}}{\mathrm{d}t}=\mathbf{M}_{pw}\mathbf{V}+\frac{J_+}{2}\mathbf{Q}_{pw}\hat{g}\mathbf{V},
\end{equation}
where $\mathbf{M}_{pw}$ is diagonal,
\begin{equation}\label{eq:KM65}
		\mathbf{M}_{pw}=
		\begin{pmatrix}
			\mathcal{X} & 0 & 0\\
			0 & \mathcal{\overline{X}} & 0 \\
			0 & 0 & \mathcal{Y}
		\end{pmatrix},
\end{equation}
and $\mathbf{Q}_{pw}$ is block diagonal,
\begin{equation}\label{eq:KM66}
	\mathbf{Q}_{pw}=
\begin{pmatrix}
	1 & -\alpha_{pw}^2 & 0 \\
	-\overline{\alpha}_{pw}^2 & 1 & 0 \\
	0 & 0 & 1
\end{pmatrix}.
\end{equation}
The auxiliary quantities are
\begin{subequations}\label{eq:KM67}
	\begin{equation}\label{eq:KM67a}
		\mathcal{X}(v,\omega)=i(v+\omega)-q\alpha_{pw},
	\end{equation}
	\begin{equation}\label{eq:KM67b}
		\mathcal{Y}(v,\omega)=i(v-\omega)-ip\text{Im}(\alpha_{pw}).
	\end{equation}
\end{subequations}

Setting $\dot{\mathbf{V}}=\lambda \mathbf{V}$ and assuming $\lambda\in\mathbb{C}\backslash \{\mathcal{X},\mathcal{\overline{X}},\mathcal{Y}\}$, we obtain
\begin{equation}\label{eq:KM68}
	\mathbf{V}=(\lambda \mathbf{I}_3 - \mathbf{M}_{pw})^{-1}\frac{J_+}{2}\mathbf{Q}_{pw}\hat{g}\mathbf{V}.
\end{equation}
Applying $\hat{g}$ to both sides and using that $\hat{g}\mathbf{V}$ is a constant vector, the solvability condition becomes
\begin{equation}\label{eq:KM69}
	\det[\mathbf{I}_3-\frac{J_+}{2}\hat{g}((\lambda \mathbf{I}_3 -\mathbf{M}_{pw})^{-1}\mathbf{Q}_{pw})]=0.
\end{equation}

Because $\mathbf{M}_{pw}$ and $\mathbf{Q}_{pw}$ are block diagonal, the eigenvalue problem decouples into two orthogonal subspaces. In the two-dimensional subspace spanned by $(a,\overline{a})\in\mathbb{C}^2$, the eigenvalues satisfy
\begin{equation}\label{eq:KM70}
	\begin{vmatrix}
		1-\frac{J_+}{2}\hat{g}(\lambda -\mathcal{X})^{-1} & \frac{J_+}{2}\hat{g}(\frac{\alpha_{pw}^2}{\lambda-\mathcal{X}}) \\
		\frac{J_+}{2}\hat{g}(\frac{\overline{\alpha}_{pw}^2}{\lambda -\mathcal{\overline{X}}}) & 1-\frac{J_+}{2}\hat{g}(\lambda-\mathcal{\overline{X}})^{-1}
	\end{vmatrix}=0.
\end{equation}
By symmetry, $\lambda$ is real. Moreover, the two diagonal elements are equal and the two off-diagonal elements are equal, so Eq.~(\ref{eq:KM70}) reduces to the pair of self-consistency relations
\begin{subequations}\label{eq:KM71}
	\begin{equation}\label{eq:KM71a}
		\frac{1}{J_+}=\frac{1}{2}\hat{g}(\frac{1+\alpha_{pw}^2}{\lambda -\mathcal{X}(v,\omega)}),
	\end{equation}
	\begin{equation}\label{eq:KM71b}
		\frac{1}{J_+}=\frac{1}{2}\hat{g}(\frac{1-\alpha_{pw}^2}{\lambda -\mathcal{X}(v,\omega)}).
	\end{equation}
\end{subequations}
In the one-dimensional subspace spanned by $b$, the eigenvalues satisfy
\begin{equation}\label{eq:KM72}
	\frac{1}{J_+}=\frac{1}{2}\hat{g}[\lambda -\mathcal{Y}(v,\omega)]^{-1}.
\end{equation}

Two remarks on the spectrum of the phase-wave state are in order. First, Eq.~(\ref{eq:KM64}) shows that the continuous spectrum corresponds to the multiplication operator $\mathbf{M}_{pw}$ and consists of three branches,
\begin{equation}\label{eq:KM73}
	\sigma_c^{pw}=\{\lambda = \mathcal{X}\}\cup \{\lambda = \mathcal{\overline{X}}\}\cup \{\lambda =\mathcal{Y}\}.
\end{equation}
By Eqs.~(\ref{eq:KM67a}) and (\ref{eq:KM67b}), $\sigma_c^{pw}$ lies on the imaginary axis and the negative real axis. As in partially synchronized states of the classical Kuramoto model, this continuous spectrum does not influence the stability. Second, for $\lambda\notin\sigma_c^{pw}$ the discrete spectrum is governed by the two independent sets of equations, Eqs.~(\ref{eq:KM71}) and (\ref{eq:KM72}). When $\alpha_{pw}=0$, we have $\mathcal{X}=i(v+\omega)$ and $\mathcal{Y}=i(v-\omega)$, and the phase-wave state reduces to the desynchronization state; Eqs.~(\ref{eq:KM71}) and (\ref{eq:KM72}) then reduce to Eq.~(\ref{eq:KM26}). For $\alpha_{pw}\neq 0$, Eqs.~(\ref{eq:KM71a}) and (\ref{eq:KM71b}) coincide with the characteristic equations for partially synchronized Kuramoto states and do affect the stability (see below). Hence, the stability of the phase-wave state is determined solely by Eq.~(\ref{eq:KM72}). We next analyze Eq.~(\ref{eq:KM72}) for $\alpha_{pw}\neq 0$ to obtain the general conditions for a stable phase-wave state.

\subsubsection{Bifurcation diagram of the phase-wave state}  

As indicated by Eq.~(\ref{eq:KM67b}), $\mathcal{Y}$ contains two contributions: $v-\omega$ is the effective frequency in the $\eta$ layer, while the second term arises from $\alpha_{pw}$. To simplify the analysis, we introduce $x=v+\omega$ and $y=v-\omega$, with inverse transformation $v=(x+y)/2$ and $\omega=(x-y)/2$. The Jacobian determinant satisfies $\left|\partial(v,\omega)/\partial(x,y)\right|=1/2$. Equation~(\ref{eq:KM72}) can then be rewritten as
\begin{equation}\label{eq:KM74}
	\frac{4}{J_+}=\int_{-\infty}^{+\infty}\int_{-\infty}^{+\infty}\frac{g(x,y)}{\lambda-if(x,y)}dxdy,
\end{equation} 
where
\begin{equation}\label{eq:KM75}
	g(x,y)=g_v(\frac{x+y}{2})g_\omega(\frac{x-y}{2}),
\end{equation} 
and
\begin{equation}\label{eq:KM76}
	f(x,y)=y-p\text{Im}(\alpha_{pw}(x)).
\end{equation}

Solving Eq.~(\ref{eq:KM74}) for $\lambda$ is generally difficult. Nevertheless, useful properties can be inferred. As in Sec.~\ref{Sec III}, symmetry implies $\lambda\in\mathbb{R}$. Stability is thus determined by the critical condition $\lambda\rightarrow 0^+$. Using
\begin{equation}\label{eq:KM77}
	\lim_{\lambda \to 0^+}\frac{1}{\lambda -if(x,y)}=\pi\delta(f(x,y)),
\end{equation}
and noting that $\lambda$ is real, the principal-value part can be dropped, yielding
\begin{equation}\label{eq:KM78}
	\frac{4}{\pi J_+}=\int_{-\infty}^{+\infty}\int_{-\infty}^{+\infty}g(x,y)\delta(f(x,y))dxdy.
\end{equation}

Introduce a smooth local transformation $(x,y)\mapsto (s,t)$ such that $s=f(x,y)$ and $t$ parametrizes the curve $f(x(t),y(t))\equiv 0$. Then the integral on the right-hand side of Eq.~(\ref{eq:KM78}) can be rewritten as
\begin{equation}\label{eq:KM79}
		\begin{aligned}
		&\int_{-\infty}^{+\infty}\int_{-\infty}^{+\infty}g(x,y)\delta(s)\left|\frac{\partial(x,y)}{\partial(s,t)}\right |dsdt \\
		&=\int_{-\infty}^{+\infty}\int_{-\infty}^{+\infty}g(x,y)\delta(s)\left|\frac{\partial(s,t)}{\partial(x,y)}\right|^{-1}dsdt \\
		&=\int_{f=0} g(x(0,t),y(0,t))\left|\mathbf{J}_0\right|^{-1}dt,
		\end{aligned}
\end{equation}
where
\begin{equation}\label{eq:KM80}
		\begin{aligned}
		\left|\mathbf{J}_0\right|
		=&\left|\frac{\partial s}{\partial x}\frac{\partial t}{\partial y} - \frac{\partial s}{\partial y}\frac{\partial t}{\partial x}\right|_{s=0}
		=\left|\boldsymbol{\nabla} f \times \boldsymbol{\tau}\right|.
	\end{aligned}
\end{equation}
Geometrically, $\boldsymbol{\tau}$ is tangent to the curve $f(x,y)=0$ and $\boldsymbol{\nabla} f$ is normal to it, so $|\boldsymbol{\nabla} f \times \boldsymbol{\tau}|=|\boldsymbol{\nabla} f||\boldsymbol{\tau}|$. Using $dt=\boldsymbol{\tau}\cdot (dx,dy)^\top$, we obtain
\begin{equation}\label{eq:KM81}
	\frac{dt}{|\boldsymbol{\tau}|}=dl,
\end{equation}
where $dl$ is the arc-length element along $f=0$. Substituting into Eq.~(\ref{eq:KM79}) gives the line-integral form
\begin{equation}\label{eq:KM82}
	\int_{f(x,y)=0} \frac{g(x,y)}{|\boldsymbol\nabla f|}dl.
\end{equation}

In particular, if
\begin{equation}\label{eq:KM83}
	f(x,y)=y-\phi(x),
\end{equation}
with $\phi(x)$ smooth, then $\boldsymbol\nabla f=(-\phi'(x),1)^\top$ and $|\boldsymbol\nabla f|=\sqrt{1+(\phi')^2}$. The curve $f=0$ is $y=\phi(x)$ and $dl=\sqrt{1+(\phi')^2}dx$, so Eq.~(\ref{eq:KM82}) reduces to $\int g(x,\phi(x))dx$. With $\phi(x)=p\text{Im}(\alpha_{pw}(x))$, Eq.~(\ref{eq:KM78}) becomes
\begin{equation}\label{eq:KM84}
	\frac{4}{\pi J_+}=\int_{-\infty}^{+\infty}g(x,\phi(x))dx.
\end{equation}

We now analyze the stability of the phase-wave state according to Eq.~(\ref{eq:KM84}). Before proceeding, we briefly discuss the qualitative structure of the stable region. For real $\lambda$, Eq.~(\ref{eq:KM72}) can be written as
\begin{equation}\label{eq:KM85}
	\frac{4}{J_+}=\int_{-\infty}^{+\infty}\int_{-\infty}^{+\infty} g(x,y)\frac{\lambda}{\lambda^2+(y-\phi(x))^2}dxdy.
\end{equation}
As in the classical Kuramoto setting, $J_+\ge 0$ implies that any solution requires $\lambda>0$; otherwise, no eigenvalue exists. This is the Landau damping scenario~\cite{83,84,85}. In infinite-dimensional settings, the discrete spectrum may be empty and convergence is weak~\cite{86,87,88,89}. Thus, in our model, the phase-wave state is either linearly unstable ($\lambda>0$) or asymptotically stable (no discrete eigenvalue).

The right-hand side of Eq.~(\ref{eq:KM85}) is a smooth, bounded function of $\lambda$. For sufficiently large $J_+$ the left-hand side becomes small, and intersections with the right-hand side may occur, leading to linear instability. Hence, stability requires sufficiently large $|J_-|$ (through $\phi$) to suppress such intersections. In particular, for $J_-=0$ or $q=0$, one can show that the phase-wave state is linearly unstable whenever $J_+>J_{+,c}$. Therefore, in the $(J_+,J_-)$ plane, the phase-wave state can only be stable for $J_+>J_{+,c}$.

To locate the stability boundary more precisely, we simplify Eq.~(\ref{eq:KM84}). Since $x$ spans the real line, the integral must be decomposed into a locked and a drifting part. From Eq.~(\ref{eq:KM67b}),
\begin{equation}\label{eq:KM86}
	\phi(x)=
	\begin{cases}
		\frac{p}{q}x, &|x|<q \\
		\frac{p}{q}x -\text{sgn}(x)p\sqrt{(\frac{x}{q})^2 -1}, &|x|>q.
	\end{cases}
\end{equation}
Letting $x=qs$, Eq.~(\ref{eq:KM84}) becomes
\begin{equation}\label{eq:KM87}
	\frac{4}{\pi q J_+}=\int_{-\infty}^{+\infty}g(qs,\phi(qs))ds.
\end{equation}
Define
\begin{equation}\label{eq:KM88}
	\varphi(s)=s-\text{sgn}(s)\sqrt{s^2 -1}.
\end{equation}
Then the stability boundary is given implicitly by
\begin{equation}\label{eq:KM89}
	\frac{4}{\pi q J_+}=\int_{|s|<1}g(qs,ps)ds +\int_{|s|>1}g(qs,p\varphi(s))ds.
\end{equation}

For the Lorentz--Lorentz case with $\Delta_v=\Delta_\omega=1$, one has $J_{+,c}=4$ and $S=\sqrt{1-4/J_+}$, hence $q=J_+S=\sqrt{J_+(J_+-4)}$ and $p=J_-S=J_-\sqrt{1-4/J_+}$. By symmetry, positive and negative $s$ contribute equally. Using the explicit forms of $g_v(v)$ and $g_\omega(\omega)$, Eq.~(\ref{eq:KM89}) reduces to
\begin{equation}\label{eq:KM90}
	\frac{2\pi}{qJ_+}=I_1 +I_2 ,
\end{equation}
with
\begin{subequations}\label{eq:KM91}
	\begin{equation}\label{eq:KM91a}
		I_1=\int_{0}^{1}\frac{1}{\frac{1}{4}(qs+ps)^2 +1}\frac{1}{\frac{1}{4}(qs-ps)^2 +1}ds,
	\end{equation}
	\text{and}
	\begin{equation}\label{eq:KM91b}
		\begin{aligned}
			I_2= \int_{1}^{+\infty}&\frac{1}{\frac{1}{4}(qs+ps-p\sqrt{s^2 -1})^2 +1}\\
			&\frac{1}{\frac{1}{4}(qs-ps+p\sqrt{s^2 -1})^2 +1}ds.
		\end{aligned}
	\end{equation}
\end{subequations}

\begin{figure}[tbp]
\begin{center}
\includegraphics[width=\linewidth]{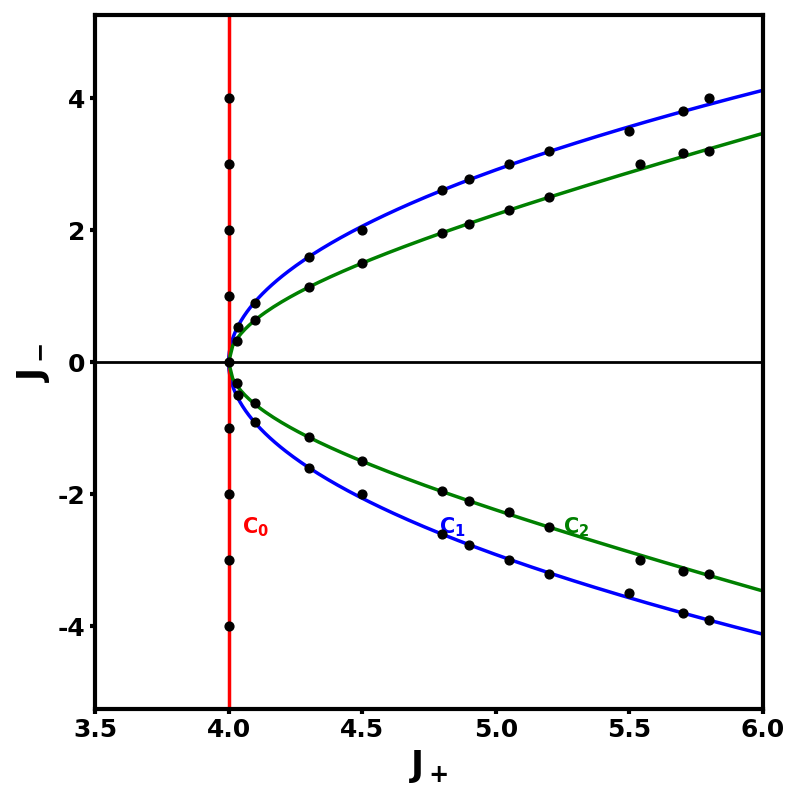}
\caption{Bifurcation diagram in the $(J_-,J_+)$ plane for the Lorentz--Lorentz case with $\Delta_v=\Delta_\omega=1$. Curves $C_0$, $C_1$, and $C_2$ denote the stability boundaries of the desynchronization, phase-wave, and synchronization states, respectively. Black dots are numerical results for $N=5\times 10^4$; solid curves are theoretical predictions.}
\vspace{-0.2cm}
\label{fig3}
\end{center}
\end{figure}

Because Eq.~(\ref{eq:KM90}) is invariant under $J_-\mapsto -J_-$, the stability boundary is symmetric about $J_-=0$. Figure~\ref{fig3} shows the curve $C_1$ obtained from Eq.~(\ref{eq:KM90}), together with numerical results for $N=5\times 10^4$, demonstrating excellent agreement.

We next explain why $C_1$ is tangent to the vertical line $C_0$ at $(J_{+,c},0)$. Let $J_+=4+\varepsilon$ and assume $J_-\sim \varepsilon^\mu$, where $\mu$ is a scaling exponent. The left-hand side of Eq.~(\ref{eq:KM90}) scales as $\sim\frac{1}{q}(\frac{\pi}{2}-\frac{\pi}{8}\varepsilon)$. In the limit $q,p\to 0$, the dominant contribution to the right-hand side is
\begin{equation}\label{eq:KM92}
	\int_{0}^{+\infty}\frac{1}{\frac{1}{4}(qs+ps)^2 +1}\frac{1}{\frac{1}{4}(qs-ps)^2 +1}ds=\frac{\pi}{2q},
\end{equation}
matching the leading term on the left-hand side and showing that $C_1$ emanates from $(4,0)$.

To determine $\mu$, we assume the next correction in Eq.~(\ref{eq:KM90}) is proportional to $(p/q)C$ with $C$ finite. Since $p\sim \varepsilon^{\frac{1}{2}+\mu}$, matching the first-order correction implies $\mu=\frac{1}{2}$. Transforming back to $(J,K)$, this yields the local parametric scaling $J\sim 4+\sqrt{\varepsilon}$ and $K\sim 4-\sqrt{\varepsilon}$, so the curve is tangent to $J+K=8$ at $(4,4)$ as $\varepsilon\to 0$, consistent with Ref.~\cite{41}.

In summary, Eq.~(\ref{eq:KM89}) provides an implicit stability boundary for the phase-wave state under general frequency--velocity distributions. In the $(J_+,J_-)$ plane, the curve $C_1$ forms a right-opening, parabola-like boundary with vertex at $(J_{+,c},0)$. For $J_+>J_{+,c}$, the phase-wave state is observable only in the region on the stable side of $C_1$. Physically, the phase-wave state is destabilized when $|J_-|$ is too small: disorder in the $\eta$ layer induced by effective-frequency heterogeneity cannot overcome the synchronization tendency generated by the strong intra-layer coupling, causing the $\eta$ layer to synchronize. This feedback then destabilizes the phase-wave state.

\subsection{The synchronization and mixed states}

Before formally analyzing the synchronization state, it is necessary to briefly comment on the method employed in this study. In general, whether for the desynchronization state and phase-wave state discussed earlier, or for the synchronization state to be examined next, our basic strategy is based on the linearization of the stationary solutions of the OA evolution equation. The corresponding linear perturbation operator can always be decomposed into a product operator plus a finite-rank perturbation operator. The spectrum of the product operator corresponds to the continuous spectrum of the system. According to the general theory of operator spectra, the continuous spectrum remains invariant under finite-rank perturbations. Such spectra typically lie entirely in the left half of the complex plane and therefore do not contribute to the instability of the state under consideration. The actual instability of these states is determined by the discrete spectrum of the linear operator, which reduces to an eigenvalue problem in linear algebra. For further details, see Ref. ~\cite{97}. Following this line of reasoning, we will conduct a detailed analysis of the stability of the synchronization state.

In the layered-network picture, the synchronization state corresponds to partial synchronization in both layers, with order parameters $(S,S)$. Unlike the desynchronization and phase-wave states, in which the layers are fully or partially decoupled, the synchronization state exhibits strong interdependence between the $\zeta$ and $\eta$ layers, making the analysis more complex.

For a steady synchronization state ($\dot{\alpha}=\dot{\beta}=0$ in Eqs.~(\ref{eq:KM54a}) and (\ref{eq:KM54b})), the first Fourier coefficients are
\begin{subequations}\label{eq:KM93}
	\begin{equation}\label{eq:KM93a}
		\alpha_s=H(\frac{qx-py}{q^2 -p^2})=H(\frac{v}{JS}+\frac{\omega}{KS}),
	\end{equation}
	\begin{equation}\label{eq:KM93b}
		\beta_s=H(\frac{qy-px}{q^2 -p^2})=H(\frac{v}{JS}-\frac{\omega}{KS}),
	\end{equation}
\end{subequations}
with $x=v+\omega$, $y=v-\omega$, and
\begin{equation}\label{eq:KM94}
	H(s)=
	\begin{cases}
		\sqrt{1-s^2}+is, &|s|<1, \\
		i(s-\text{sgn}(s)\sqrt{s^2 -1}), &|s|>1.
	\end{cases}
\end{equation}

Substituting Eqs.~(\ref{eq:KM93a}) and (\ref{eq:KM93b}) into Eqs.~(\ref{eq:KM55a}) and (\ref{eq:KM55b}) yields a self-consistent equation for $S$ under general frequency--velocity distributions. This result differs from the classical Kuramoto case only through the distribution of the effective frequencies. By mirror symmetry, $\alpha_s$ and $\beta_s$ have analogous forms, and the resulting order parameters are identical.

For the Lorentz--Lorentz case with $\Delta_v=\Delta_\omega=1$, one obtains
\begin{equation}\label{eq:KM95}
	S=\sqrt{1-\frac{4J_+}{J_+^2 -J_-^2}}.
\end{equation}
This expression fully agrees with the result obtained in Ref. \cite{41}, where it was validated through numerical analysis. The corresponding critical curve is
\begin{equation}\label{eq:KM96}
	J_-=\pm \sqrt{J_+(J_+ -4)},
\end{equation}
denoted as $C_2$ in Fig.~\ref{fig3}. The synchronization state exists on the right side of $C_2$. The curve $C_2$ is a right-opening parabola with vertex at $(4,0)$, and $S=0$ on $C_2$. For fixed $J_+$, $S$ increases with distance from $C_2$, and attains its maximal value at $J_-=0$.

Compared to the phase-wave state, the parameter region supporting the synchronization state is narrower. When $|J_-|$ is small, the coupling between layers is weak and each layer tends to evolve nearly independently; mirror symmetry then drives comparable synchronization levels in both layers. Moreover, for small $|J_-|$, the effective frequencies in the two layers become closer, reducing the disorder that could counterbalance the intra-layer synchronization tendency induced by $J_+$. As a consequence, smaller $|J_-|$ typically yields larger $S$ at fixed $J_+$.

The stability of the synchronization state can be studied by perturbing the steady state in Eqs.~(\ref{eq:KM93a}) and (\ref{eq:KM93b}). Proceeding as for the desynchronization and phase-wave states yields
\begin{equation}\label{eq:KM97}
	\frac{\mathrm{d}\mathbf{V}}{\mathrm{d}t}=\mathbf{M}_s \mathbf{V}+\mathbf{Q}_s \hat{g}\mathbf{V},
\end{equation}
with
\begin{equation}\label{eq:KM98}
	\mathbf{M}_s=
	\begin{pmatrix}
		c_1 & 0 & c_2 & -c_2 \\
		0 & \overline{c}_1 & -\overline{c}_2 &\overline{c}_2 \\
		d_2 & -d_2 & d_1 & 0\\
		-\overline{d}_2 & \overline{d}_2 & 0 &\overline{d}_1
	\end{pmatrix},
\end{equation}
and
\begin{equation}\label{eq:KM99}
	\mathbf{Q}_s=
	\begin{pmatrix}
		c_3 & c_4 &c_5 & c_6 \\
		\overline{c}_4 & \overline{c}_3 & \overline{c}_6 & \overline{c}_5 \\
		d_5 & d_6 & d_3 & d_4 \\
		\overline{d}_6 & \overline{d}_5 & \overline{d}_4 & \overline{d}_3
	\end{pmatrix}.
\end{equation}
The entries are
\begin{subequations}\label{eq:KM100}
	\begin{equation}\label{eq:KM100a}
		c_1=ix-q\alpha_s +\frac{p}{2}(\overline{\beta}_s -\beta_s),
	\end{equation}
	\begin{equation}\label{eq:KM100b}
		d_1=iy-q\beta_s+\frac{p}{2}(\overline{\alpha}_s -\alpha_s),
	\end{equation}
	\begin{equation}\label{eq:KM100c}
		c_2=-\frac{p}{2}\alpha_s,
	\end{equation}
	\begin{equation}\label{eq:KM100d}
		d_2=-\frac{p}{2}\beta_s,
	\end{equation}
	\begin{equation}\label{eq:KM100e}
		c_3=d_3=\frac{J_+}{2},
	\end{equation}
	\begin{equation}\label{eq:KM100f}
		c_4=-\frac{J_+}{2}\alpha_s^2,
	\end{equation}
	\begin{equation}\label{eq:KM100g}
		d_4=-\frac{J_+}{2}\beta_s^2,
	\end{equation}
	\begin{equation}\label{eq:KM100h}
		c_5=\overline{d}_5=\frac{J_-}{2}\alpha_s \overline{\beta}_s,
	\end{equation}
	\begin{equation}\label{eq:KM100i}
		c_6=d_6=-\frac{J_-}{2}\alpha_s \beta_s.
	\end{equation}
\end{subequations}

The continuous spectrum is determined by the eigenvalues of $\mathbf{M}_s$. Using a Schur-complement calculation for this block matrix yields the characteristic polynomial
\begin{equation}\label{eq:KM101}
	\sum_{i=0}^{4}a_i \lambda^i =0,
\end{equation}
with coefficients
\begin{subequations}\label{eq:KM102}
	\begin{equation}\label{eq:KM102a}
		a_4=1,
	\end{equation}
	\begin{equation}\label{eq:KM102b}
		a_3=-2\text{Re}(c_1+d_1),
	\end{equation}
	\begin{equation}\label{eq:KM102c}
		a_2=|c_1|^2+|d_1|^2 +(c_1 +\overline{c}_1)(d_1 +\overline{d}_1)-(c_2 +\overline{c}_2)(d_2 +\overline{d}_2),
	\end{equation}
	\begin{equation}\label{eq:KM102d}
		\begin{aligned}
		a_1=&(d_1 \overline{d}_2 +\overline{d}_1 d_2)(c_2 +\overline{c}_2)+(c_1 \overline{c}_2 +\overline{c}_1 c_2)(d_2 +\overline{d}_2)\\
		&-|d_1|^2(c_1 +\overline{c}_1)-|c_1|^2(d_1 +\overline{d}_1),
		\end{aligned}
	\end{equation}
	\begin{equation}\label{eq:KM102e}
		a_0 =|c_1|^2|d_1|^2 -(c_1\overline{c}_2 +\overline{c}_1 c_2)(d_1 \overline{d}_2 +\overline{d}_1 d_2).
	\end{equation}
\end{subequations}

\begin{figure}[tbp]
\begin{center}
\includegraphics[width=\linewidth]{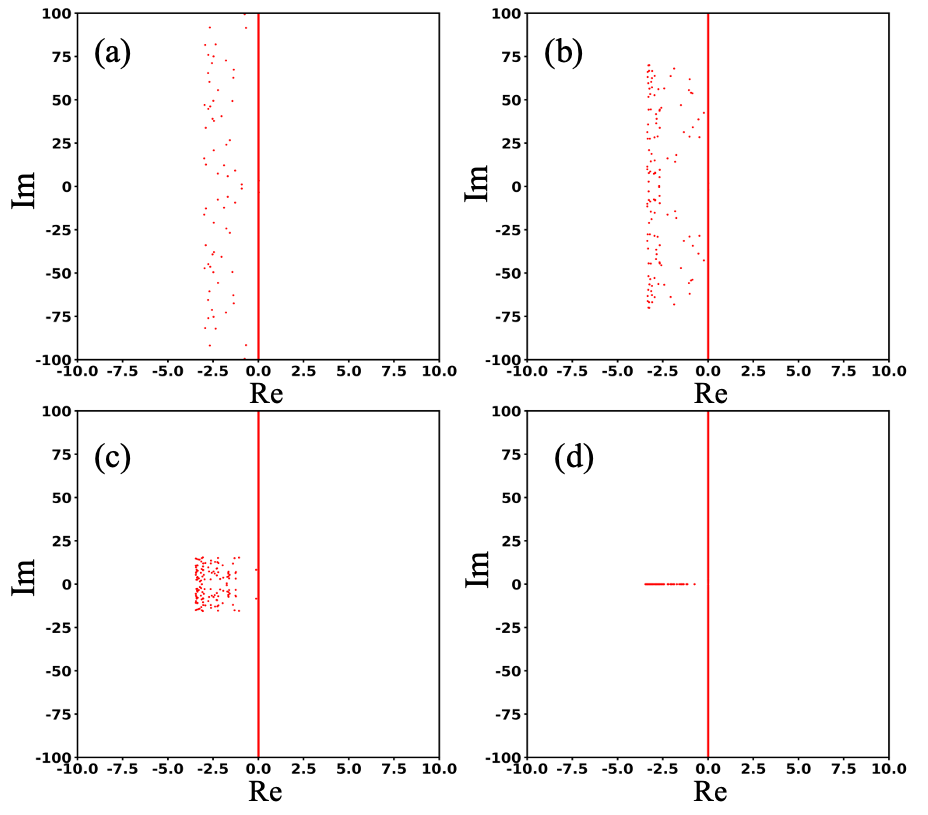}
\caption{Continuous spectrum of the synchronization state for Lorentz--Lorentz distributions with $\Delta_v=\Delta_\omega=1$. The parameter sets $(J_+,J_-)$ are (a) $(6,2)$, (b) $(6,1)$, (c) $(6,0.2)$, and (d) $(6,0)$, all inside the region bounded by $C_2$ in Fig.~\ref{fig3}. As $J_-\to 0$, the spectrum approaches the union of the negative real axis and the imaginary axis, as in partially synchronized states of the classical Kuramoto model.}
\vspace{-0.2cm}
\label{fig4}
\end{center}
\end{figure}

Since Eq.~(\ref{eq:KM101}) does not yield an explicit closed-form expression for $\lambda$, we use numerical computations to illustrate the continuous spectrum. Fixing $J_+=6$, Fig.~\ref{fig4} shows the distribution of $\lambda$ in the complex plane for several values of $J_-$ (all within the synchronization region) for the Lorentz--Lorentz case. In all cases, the continuous spectrum lies in the left half-plane (i.e., $\text{Re}(\lambda)\le 0$), and therefore does not destabilize the synchronization state. In particular, as $|J_-|\to 0$ [Fig.~\ref{fig4}(d)], the spectrum approaches the negative real axis and the imaginary axis, similar to partially synchronized states in the classical Kuramoto model.

The discrete spectrum is determined by
\begin{equation}\label{eq:KM103}
	\det(\mathbf{I}_4 -\hat{g}[(\lambda\mathbf{I}_4 -\mathbf{M}_s)^{-1} \mathbf{Q}_s])=0.
\end{equation}
Although Eq.~(\ref{eq:KM103}) determines the discrete eigenvalues in principle, obtaining them analytically (or even numerically) is difficult, because the characteristic condition is a nonlinear integral equation in $\lambda$ and, for general distributions, $S$ itself lacks a closed form.

Motivated by the classical Kuramoto model and by the spectral structures of the desynchronization and phase-wave states, we conjecture that the discrete spectrum of the synchronization state given by Eq.~(\ref{eq:KM103}) is empty, i.e., that the synchronization state exhibits Landau damping. A rigorous proof is beyond the scope of this work. Below, we provide supporting evidence by considering the special case $ J _ {-} = 0$ (i.e., $p=0$), for which the two layers decouple and each reduces to a classical Kuramoto model with effective frequencies. In this case, $\mathbf{M}_s$ and $\mathbf{Q}_s$ decompose into the direct sum of two subblocks corresponding to the $\zeta$ and $\eta$ layers. Without loss of generality, we focus on the $\zeta$ layer.

Restricting Eq.~(\ref{eq:KM103}) to this subspace yields
\begin{subequations}\label{eq:KM104}
	\begin{equation}\label{eq:KM104a}
		\frac{1}{J_+}=h_c(\lambda)=\frac{1}{2}\hat{g}\frac{1-\alpha_s^2}{\lambda-c_1},
	\end{equation}
	\begin{equation}\label{eq:KM104b}
		\frac{1}{J_+}=h_s(\lambda)=\frac{1}{2}\hat{g}\frac{1+\alpha_s^2}{\lambda-c_1},
	\end{equation}
\end{subequations}
where $c_1$ corresponds to the continuous spectrum and is given by
\begin{equation}\label{eq:KM105}
	c_1=
	\begin{cases}
		-\sqrt{q^2 -x^2}, &|x|<q, \\
		i\text{sgn}(x)\sqrt{x^2 -q^2}, &|x|>q.
	\end{cases}
\end{equation}
To assess whether Eqs.~(\ref{eq:KM104a}) and (\ref{eq:KM104b}) admit nontrivial roots, we introduce $s=x/q$ and rescale $\lambda\rightarrow \lambda/q$. Substituting $\alpha_s$ and $c_1$ into Eq.~(\ref{eq:KM104a}), using the self-consistency relation $J_+^{-1}=\int_{-1}^{1}\sqrt{1-s^2}\tilde{g}(qs)ds$ and symmetry, we obtain
\begin{equation}\label{eq:KM106}
	\begin{aligned}
		\Delta_1(\lambda)& = \frac{1}{2}(h_c(\lambda)-J_+^{-1})\\
		& =\int_{0}^{1}I_1(\lambda)\tilde{g}(qs)ds +\int_{1}^{+\infty}I_2(\lambda)\tilde{g}(qs)ds,
	\end{aligned}
\end{equation}
with
\begin{subequations}\label{eq:KM107}
	\begin{equation}\label{eq:KM107a}
		I_1(\lambda)=\frac{s^2}{\lambda +\sqrt{1-s^2}}-\sqrt{1-s^2},
	\end{equation}
	\text{and}
	\begin{equation}\label{eq:KM107b}
		I_2(\lambda)=\frac{\lambda s^2 (1-\sqrt{1-s^{-2}})}{\lambda^2 +s^2 -1}.
	\end{equation}
\end{subequations}
One can show that for $\lambda\ge 0$ the quantity $\Delta_1(\lambda)$ is bounded above by a nonnegative constant.

The sign of $I_1(\lambda)$ changes over $s\in(0,1)$: there exists $s_0\in(0,1)$ such that $I_1(\lambda)\le 0$ for $s\le s_0$ and $I_1(\lambda)>0$ for $s>s_0$. Since $\tilde{g}(x)$ is unimodal and symmetric, it is monotonically decreasing for $x>0$, implying
\begin{equation}\label{eq:KM108}
	\int_{0}^{1}I_1(\lambda)\tilde{g}(qs)ds \le \tilde{g}(qs_0)\int_{0}^{1}I_1(\lambda)ds.
\end{equation}
Moreover, as $I_2(\lambda)\ge 0$ for $\lambda\ge 0$, we have
\begin{equation}\label{eq:KM109}
	\int_{1}^{+\infty}I_2(\lambda)\tilde{g}(qs)ds \le \tilde{g}(q)\int_{1}^{+\infty}I_2(\lambda)ds.
\end{equation}
Since $I_1(\lambda)$ decreases monotonically in $\lambda$ and $\int_{0}^{1}I_1(0)ds=0$, we also have
\begin{equation}\label{eq:KM110}
	\int_{0}^{1}I_1(\lambda)\tilde{g}(qs)ds \le \tilde{g}(q)\int_{0}^{1}I_1(\lambda)ds.
\end{equation}
Combining these estimates yields
\begin{equation}\label{eq:KM111}
	\Delta_1(\lambda) \le \tilde{g}(q)\left[\int_{0}^{1}I_1(\lambda)ds+\int_{1}^{+\infty}I_2(\lambda)ds\right] 
	=\tilde{g}(q)\frac{\pi}{4}.
\end{equation}
A similar argument shows
\begin{equation}\label{eq:KM112}
	\Delta_2(\lambda)=\frac{1}{2}(h_s(\lambda)-J_+^{-1}) \le 0.
\end{equation}
Therefore, apart from the trivial root $\lambda=0$ arising from rotational invariance, no discrete eigenvalue exists in this decoupled case. This supports the conjecture that typical synchronization states have an empty discrete spectrum and are at most asymptotically stable.

Finally, we would like to give some comments on the mechanism and properties of the intermediate mixed state. As a state generated by symmetry-breaking, the mixed state typically appears in the transition from the phase-wave state to the synchronization state, and is characterized by $(S_1,S_2)$, with $S_1\neq S_2$. Mirror symmetry implies that the states $(S_1,S_2)$ and $(S_2,S_1)$ occur with equal probability. For a steady mixed state ($\dot{\alpha}=\dot{\beta}=0$), one has
\begin{subequations}\label{eq:KM113}
	\begin{equation}\label{eq:KM113a}
		\alpha_m=H\left(\frac{\Omega_1}{S_1 JK} \right) ,
	\end{equation}
	\begin{equation}\label{eq:KM113b}
		\beta_m=H\left(\frac{\Omega_2}{S_2 JK} \right) ,
	\end{equation}
\end{subequations}
where $\Omega_{1,2}=Kv\pm J\omega$ and $H$ is defined in Eq.~(\ref{eq:KM94}).

Equations~(\ref{eq:KM113a}) and (\ref{eq:KM113b}) show that the mixed state can be viewed as two Kuramoto-type populations with effective frequencies $\Omega_1$ and $\Omega_2$ and coupling strength $JK$. Although $\Omega_1$ and $\Omega_2$ have the same marginal distribution, they are correlated; consequently, the phase-locking conditions $|\Omega_{1,2}|\le JKS_{1,2}$ need not yield $S_1=S_2$, providing a microscopic origin for the emergence of the mixed state. The special case $S_1=S_2$ corresponds to the synchronization state. Since the phase-locking intervals cannot be obtained analytically in general, explicit self-consistency relations for $(S_1,S_2)$ are difficult to derive. Instead, one can expect that the mixed state occupies the region between the phase-wave boundary $C_1$ and the synchronization boundary $C_2$ in the parameter space. That is, as the coupling increases, the system typically transitions from the phase-wave to mixed and then to synchronization states.

Overall, our results suggest that the synchronization state arises from the interplay between the two layers for sufficiently small $|J_-|$, where quenched disorder within each layer is not too strong and coherent motion can develop in both layers. The mixed state, by contrast, emerges through symmetry breaking associated with the instabilities of the phase-wave and synchronization states. In particular, the codimension-two point $(J_{+,c},0)$ marks the intersection of the stability boundaries $C_0$ and $C_1$ in Fig.~\ref{fig3}, and rich bifurcation behavior occurs in its vicinity in the parameter space $(J_+,J_-)$.

\section{Conclusion}\label{Sec VI}

To summarize, we have revisited the one-dimensional ring-structured swarmalator model and investigated, theoretically and numerically, the dynamical properties of the collective states observed in the system. From the perspective of layered complex networks, we have uncovered an intrinsic connection between the one-dimensional swarmalator model and the high-dimensional agent model studied in previous works, and it is found that the collective dynamics in these seemingly different settings are governed by the same low-dimensional invariant manifold, namely the generalized Ott--Antonsen manifold. Leveraging this connection, we have analyzed four representative collective states in the one-dimensional model under general frequency and velocity distributions: the desynchronization state, the phase-wave state, the synchronization state, and the mixed state. Our main contributions are as follows.

\begin{enumerate}
\item {\textit{Stability of the desynchronization state.}}
By mapping the system to a two-layer Kuramoto description and incorporating the probability density in each layer, we have derived the self-consistent equations for the eigenvalues of the linear operator associated with the desynchronization state, and clarified the corresponding spectral structures. This yielded a general expression for the critical coupling strength that determines the stability of the desynchronization state [Eq.~(\ref{eq:KM35})]. Notably, this criterion is a direct generalization of the classical Kuramoto result and does not depend on the detailed form of the frequency--velocity convolution distribution; instead, the threshold is controlled by the value of the convolution density at the origin.

\item {\textit{Bridge to high-dimensional agent dynamics.}}
We have demonstrated that Tanaka's high-dimensional agent model provides a natural bridge between the microscopic high-dimensional description and the reduced swarmalator dynamics. In particular, the generalized Möbius transformation (equivalently, the generalized Ott--Antonsen manifold) offers a rigorous basis for dimension reduction in this class of systems, and supplies a unifying theoretical viewpoint for spatiotemporal self-organization and swarming behaviors in active matter systems.

\item {\textit{Stability of the phase-wave state.}}
For general frequency--velocity distributions, we have derived an implicit integral equation for the critical condition of the phase-wave state. This analysis reveals that the stability of the phase-wave state is primarily controlled by the inter-layer coupling strength. When the inter-layer coupling is too weak, disorder in the $\eta$ layer induced by effective-frequency heterogeneity cannot counterbalance the synchronization tendency driven by strong intra-layer coupling. As a result, the incoherent state of the $\eta$ layer loses stability, and the ensuing feedback destabilizes the phase-wave state through its coupling to the $\zeta$ layer.

\item {\textit{Mechanisms for the synchronization and mixed states.}}
We have clarified the mechanisms underlying the synchronization and mixed states. The synchronization state emerges from the interaction between the two layers and is stable when the inter-layer coupling magnitude $|J_-|$ is sufficiently small, so that quenched disorder within each layer is not excessively strong and macroscopic coherence can develop in both layers. The mixed state, by contrast, arises through symmetry breaking when either the phase-wave state or the synchronization state becomes unstable. In the $(J_+,J_-)$ parameter plane, the three critical curves (for the desynchronization, phase-wave, and synchronization states) intersect at the critical point $(J_{+,c},0)$, around which the system exhibits rich bifurcation behavior.
\end{enumerate}

Finally, we note that the theoretical framework developed in the present work is not limited to the pairwise-coupled ring model. It can be extended to more general swarmalator systems, including those with higher-order interactions, phase shifts or spatial frustration (e.g., Kuramoto--Sakaguchi-type couplings), time-delayed or weighted interactions, and external force~\cite{98,99}. Such extensions may provide further insight into the mechanisms and universality of collective behaviors in swarmalator systems, and, more broadly, into pattern formation in active matter systems. Several promising research topics warrant further investigation along the same direction are the following. First, the role of Landau damping in shaping the stability properties of the desynchronization, phase-wave, and synchronization states remains unclear. Second, a systematic bifurcation analysis in the neighborhood of the codimension-two critical point is still lacking; in particular, explicit criteria for non-codimension-one bifurcations would help reveal universal features of the transitions near $(J_{+,c},0)$~\cite{90,91,92}. Finally, applying the generalized Ott--Antonsen manifold to higher-dimensional swarmalator models may uncover new spatiotemporal phenomena~\cite{93,94}. The present work provides a theoretical foundation for pursuing these topics.

\section*{ACKNOWLEDGMENTS}
This work is supported by the National Natural Science Foundation of China (Grant Nos.11905068 and 12275165) and the Natural Science Foundation of Fujian Province, China (Grant No.~2023J01113).

\end{document}